\documentclass[a4paper, notitlepage,12pt]{article}
\usepackage{amsmath}
\usepackage[polutonikogreek,english]{babel}
\usepackage[or]{teubner}
\usepackage{amssymb}
\usepackage{eurosym}
\usepackage{graphicx}
\usepackage{fancyhdr}
\usepackage[a4paper, left=1.8in, right=1.3in]{geometry}
\usepackage{array}
\usepackage{epsfig}
\usepackage{subfigure}
\usepackage{amsmath}
\usepackage{geometry}
\usepackage{natbib}
\usepackage{mathrsfs}
\usepackage{rotate}
\usepackage{lscape}
\usepackage{setspace}
\usepackage{amssymb}
\usepackage{amsfonts}
\usepackage{amsmath}
\usepackage{setspace}
\usepackage{tabularx}
\usepackage{lscape}
\usepackage{rotating}
\usepackage{hyperref}
\usepackage{xtab}
\usepackage{chngpage}
\usepackage{pdflscape}
\usepackage{rotfloat}
\usepackage{color}
\usepackage{booktabs}
\usepackage{anysize}
\usepackage{longtable}
\usepackage{apalike}
\usepackage{animate}
\usepackage{placeins}
\usepackage{siunitx}
\usepackage{dcolumn}

\usepackage{booktabs}
\allowdisplaybreaks
\newtheorem{theorem}{Theorem}[section]

\newtheorem{corollary}{Corollary}

\renewcommand{\theequation}{\thesection.\arabic{equation}}

\newcounter{rmk}
\newenvironment{rmk}[1][]{\refstepcounter{rmk}\par\medskip\noindent
\textbf{Remark~\thermk.\ #1} \rmfamily} {\smallskip}
\newenvironment{remark}{\smallskip \begin{rmk}}{\hfill $\diamondsuit$ \end{rmk}  }
\numberwithin{equation}{section} 

\begin{document}

\title{Measuring wage inequality under right censoring\thanks{We thank Daron Acemoglu and participants of the Workshop on \textit{New Frontiers in Statistics of Extremes} and of the $IX_t$ Workshop in \textit{Time Series Econometrics} for helpful comments and suggestions, and Filipe Caires for excellent research assistance. This work was funded by Fundação para
a Ciência e Tecnologia through project number UID/Multi/00491/2019, PTDC/EGE-ECO/28924/2017, UID/ECO/00124/2013,
UID/ECO/00124/2019, UID/GES/00407/2013 and Social Sciences DataLab (LISBOA-01-0145-FEDER-022209), POR Lisboa (LISBOA-01-0145-FEDER-007722,
LISBOA-01-0145-FEDER-022209), and POR Norte (LISBOA-01-0145-FEDER-022209).}}
\author{Jo\~{a}o Nicolau$^{a}$, Pedro Raposo$^{b}$ and Paulo M. M.
Rodrigues$^{c}$ \\
$^{a}$ {\small {}ISEG-Universidade de Lisboa and CEMAPRE}\\
$^{b}$ {\small {} Cat\'olica Lisbon School of Business and Economics}\\
$^{c}$ {\small {}Banco de Portugal and Nova School of Business and Economics}%
}

\date{\today}
\maketitle

\begin{abstract}
\noindent In this paper we investigate potential changes which may have occurred over the last two decades in the probability mass of the right tail of the wage distribution,  through the analysis of the corresponding tail index. In specific, a conditional tail index estimator is introduced which explicitly allows for right tail censoring (top-coding), which is a feature of the widely used current population survey (CPS), as well as of other surveys. Ignoring the top-coding may lead to inconsistent estimates of the tail index and to under or over statements of inequality and of its evolution over time. Thus, having a tail index estimator that explicitly accounts for this sample characteristic is of importance to better understand and compute the tail index dynamics in the censored right tail of the wage  distribution. The contribution of this paper is threefold: i) we introduce a conditional tail index estimator that explicitly handles the top-coding problem, and evaluate its finite sample performance and compare it with competing methods; ii) we highlight that the factor values used to adjust the top-coded wage have changed over time and depend on the characteristics of individuals, occupations and industries, and propose suitable values; and iii) we provide an in-depth empirical analysis of the dynamics of the US wage distribution's right tail using the public-use CPS database from 1992 to 2017.
\bigskip


\end{abstract}

\thispagestyle{empty}

\newpage

\onehalfspace

\bigskip
\section{Introduction}

The sharp rise in overall wage inequality in the second half of the 20th century has become a stylized fact (\citet{Autor(2019)} and \citet{Goosetal(2014)}). Wage inequality growth in the 1980s was followed by a slowdown in the 1990s as a result of divergent trends in the bottom and top of the wage distribution. Both the 90/50 and 50/10 indexes grew rapidly in the early 1980s, and although lower tail inequality virtually stopped growing after 1987 upper-tail inequality kept rising. The deceleration in inequality growth observed in the 1990s resulted mainly from polarization, i.e., from an abrupt stop or reversal of inequality growth in the lower-tail coupled with a sustained secular rise of the upper-tail inequality. According to \citet{Autoretal(2008)} between 1963 and 2005 the 90th percentile wage rose by more than 55\% relatively to the 10th percentile for both men and women. 	


Existing empirical evidence suggests that the rise in wage inequality is largely explained by shifts in the supply and demand for skills (\citet{GoosManning(2007)}), and by the erosion of labour market institutions (e.g. unions and minimum wage) (\cite{Kalleberg2011}). It is documented that the increase in inequality in the 1980s was the result of a secular rise in the demand for skill which faced an abrupt slowdown in the relative supply of high-skilled workers (college or equivalent) in the form of lower attainment and of smaller labor-entering cohorts which originated expanding wage differentials (\citet{Autoretal(2008)}; \citet{KatzMurphy(1992)}; \citet{CardDiNardo(2002)}; and \citet{AcemogluAutor(2011)}).

The monotonic increase of inequality until the late 1980s followed by the divergent evolution in the top and lower half of the distribution is robust to different measures and samples.\footnote{The result holds for male and female samples separately, considering weekly wages of full-time workers as well as for the March CPS samples (\citet{Autoretal(2006)}).} Steady growth in the upper-tail inequality can also be seen from the rising share of wages paid to the top 10\% and 1\% earners (\citet{PikettySaez(2003)}). However, literature based on public-use CPS data has produced  a less than perfect picture of the right tail of the wage distribution because of the top-coding (\citet{Armouretal(2016)}). 																			

The CPS wage data has historically been censored at the top (top-coded) and ignoring this fact or not adequately handling it may result in inconsistent tail index estimates, lead to understatements of inequality and affect the estimates of its dynamics (\citet{Fengetal(2006)}).\footnote{\citet{Parker(1999)} developed a model of wages in which wages follow a Generalized Beta Distribution of the second kind (GB2). \citet{Bordleyetal(1996)} show that GB2 exhibits better fit to US wage data than alternative distributions. Because the authors are modelling the distribution of total wage and not its components, they only need to know whether total wages are censored or not, and therefore do not need to be concerned with consistency problems in categories as in \citet{Burkhauseretal(2004)}. } In addition, top-coding has changed over time. For instance, the  top-coded wage was set at $\$$1923 in 1997 and changed to $\$$2884 from 1998 onward. But even during periods of constant nominal top-coding the data may hide changes in inequality (\citet{LevyMurnane(1992)}).

While some authors have tried to address the top-coding issues by restricting the sample under analysis, the method presented in this article makes use of the complete set of information available from the public use CPS data, for every year, in a time-consistent fashion, arguably providing better estimates on the level of wage inequality than other available measures. In specific, a conditional tail index regression specifically designed to account for right censoring is used. The tail index (sometimes also referred to as Pareto coefficient) is an important indicator, as it can be interpreted as an inverse measure of concentration of top wages. The lower the value of the index the more concentrated the distribution is.  

Several estimation approaches have recently been proposed which consider either non-random or random covariates; see e.g. \citet{MaJiangHuang2019} (and references therein). Our contribution falls into the latter class and provides a tail index estimator which takes the top-coding explicitly into consideration, providing in this way more efficient and consistent estimates than methods currently available in the literature.
The superior performance of the new approach is illustrated, and it is shown using the public-use CPS database from 1992 to 2017 that the factor values used for the adjustment of the top-coded wages changed over time and across the characteristics of individuals, occupations and industries; moreover it is also shown using the new estimator that the tail index has been decreasing since 1992 suggesting increased concentration in the right tail.  

The contribution of this paper is threefold: i) we introduce a conditional tail index estimator that explicitly handles the top-coding problem, evaluate its finite sample performance and compare it to competing methods; ii) we show that the factor values used for the adjustment of top-coded wages change over time and across the characteristics of individuals, occupations and industries, and suitable values are proposed; and iii) we provide an in-depth analysis of the dynamics of the US wage distribution's right tail using the public-use CPS database from 1992 to 2017.

The remainder of the paper is organized as follows. Section  2 introduces the methodology of analysis, the new tail index estimator and a detailed description of the computation of the partial effects; Section 3 presents the results of an in-depth Monte Carlo analysis on the finite sample properties of the new approach and a comparison to existing procedures; Section 4 describes and  discusses the results of an empirical analysis of the right tail characteristics of the wage distribution and wage inequality in the US using the CPS database from 1992 to 2017; and finally, in Section 5 presents the main conclusions of the paper. A technical Appendix collects proofs of the results put forward throughout the paper.

\section{Methodology}

To reduce the top-coding bias, researchers interested in measuring long-term trends in wages typically impute top-code values to create a consistent series. Until recently one of four approaches has in general been adopted in the literature: (1) the top-coding problem is ignored i.e., top-coded observations are dropped  (see e.g. \cite{JensenShore(2015)}); (2) an \textit{ad hoc} adjustment of the top-coded wages is made (e.g. \citet{Lemieux(2006)} multiplied top-coded hourly wages by 1.4, and \citet{Autoretal(2008)} multiplied top-coded weekly wages by 1.5); (3) a Pareto distribution is used to estimate wages at the top of the distribution (e.g. \citet{BernsteinMishel(1997)}, \citet{PikettySaez(2003)}); and (4) cell means or rank-proximity swapped data based on the still-censored internal CPS data is used (e.g. \cite{Larrimoreetal(2008)} and  Burkhauser et al., 2008); for a discussion and shortcomings of these approaches see, \textit{inter alia}, Burkhauser et al. (2010) and \citet{Armouretal(2016)}. 

In a recent contribution \citet{Armouretal(2016)} proposed an alternative approach which consists in estimating the tail index of a censored Pareto distribution. To briefly illustrate the procedure consider first the survival function, $\overline{F}$, of a Pareto distribution\footnote{This distribution was used, for instance, by \cite{Harrisonl(1981)} to analyse earnings by size in the UK. }
\begin{equation}
\overline{F}(y):=P\big(Y_i > y\big)=\bigg(\frac{y_0}{y_i}\bigg)^{\alpha}, \text{ where } y_i \ge y_0 > 0 \text{ and } \alpha > 0 \label{1}
\end{equation}%
and the corresponding density function, $f_Y(y)=(\alpha y_0^{\alpha})/(y_i^{\alpha+1}). $
A large number of tail index estimation procedures is available in the literature. One widely used approach is the conditional maximum likelihood estimator (MLE) proposed by \cite{Hill(1975)},%
\begin{equation}
\widehat{\alpha }_{Hill}:=\left[ \frac{1}{m}\sum \limits_{j=1}^{m}\log
y_{\left( j\right) }-\log y_{\left( 0\right) }\right] ^{-1}  \label{Hill}
\end{equation}%
where $m$ is the number of largest order statistics used in the estimation of $\alpha$, $y_{(j)}, \text{ } j=1,...,m$ are the largest $m$ order statistics and $y_0$ is the tail cut off point. 

However, recognizing the limitations of this approach when the data is top-coded, \cite{Armouretal(2016)} proposed an alternative method, which consists of an adaptation of the Hill estimator taking into consideration the censoring. This approach provides an unbiased estimate of the censored Pareto parameter, $\alpha$, while using all available information. In specific, in the case of a censored sample the outcome variable is,
\begin{equation}
\omega _{i}=\left \{ 
\begin{array}{ccc}
y_{i} & if & y_{0} \leq y_{i}<y_{c} \\ 
y_{c} & if & y_{i} \geq y_{c}%
\end{array},%
\right.  \label{wi_a}
\end{equation}
where $y_0$ is the tail cut off point and $y_c$ the top-coded value. Hence, the density function of the censored Pareto distribution is,
\begin{equation}
g_Y(\omega_i)=\bigg(\frac{\alpha y_0^{\alpha}}{\omega_i^{\alpha+1}}\bigg)^{I_{(y_0 \le \omega_i < y_c)}}\bigg[\bigg(\frac{y_0}{y_c}\bigg)^{\alpha}\bigg]^{I_{(\omega_i \ge y_c)}} \label{4}
\end{equation}%
and the respective log-likelihood function,
\begin{eqnarray}
log L &=& \sum_{i=1}^{m}g_Y(\omega_i) \notag\\
&=&\sum_{i=1}^{m}I_{(y_0 \le \omega_i < y_c)}\bigg(log(\alpha)+\alpha log(y_0) - (\alpha+1)log(\omega_i)\bigg) + \notag\\
&& \sum_{i=1}^{m}I_{(\omega_i \ge y_c)}\bigg(\alpha log(y_0) - \alpha log(y_c)\bigg). \label{2.5} 
\end{eqnarray}

Consequently, the conditional MLE estimator proposed by \cite{Armouretal(2016)} computed from (\ref{2.5}) is,
\begin{equation}
\hat{\alpha}^c_{Hill}=\frac{n_0}{\sum_{i=1}^{m}I_{(y_0 \le y_i < y_c)}log(y_i) + n_clog (y_c)-(n_0+n_c) log(y_0)} \label{5} 
\end{equation}%
where $n_0$ is the number of individuals with wages between $y_0$ and $y_c$, $n_c$ is the number of individuals with wages at or above $y_c$, and $n_0$ + $n_c$ = $m$.


\subsection{The conditional tail index estimator and properties}

In this paper, a new approach, also designed to overcome the top-coding bias, is proposed. In specific, a conditional tail index estimator which explicitly takes the right censoring of the data into account and uses covariates in the estimation process is introduced. Correctly estimating this tail index is of importance as it is  used, for instance, for the imputation of wages above the top-code. Furthermore, the procedure has the additional advantage of allowing for an in-depth analysis of the determinants that impact the tail index the strongest according to the characteristics of individuals, occupations and industries and  whether these impacts have changed over time. The use of different scaling factors to impute wages depending on the different categorizations of individuals has been used previously in the literature, see e.g.,  \cite{MacphersonHirsch(1995)} who allow the scaling factors to vary according to gender and over the years.

To introduce the conditional tail index estimation approach we consider observations ($\mathbf{X}_{i},$ $Y_{i}$), where $%
Y_{i}\in \mathbb{R}^{1}$ is the response of interest, and $\mathbf{X}%
_{i}:=(x_{1i},...,x_{pi})^{\prime }\in \mathbb{R}^{p}$ is an associated p-dimensional vector of predictors with $1\leq i\leq n$. In addition, let $F(y|\mathbf{x}; \boldsymbol{\theta}):=P[Y_{i}\leq y|\mathbf{X}_{i}=\mathbf{x}]$ be the cumulative distribution
function of $Y_{i}$ conditional on $\mathbf{X}_{i}$, and assume that the corresponding survival function (under no censoring) is,
\begin{equation}
\overline{F}(y|\mathbf{x}; \boldsymbol{\theta}):=1-F(y|\mathbf{x}; \boldsymbol{\theta})=y^{-\alpha (\mathbf{x})}\mathcal{L}(y;\mathbf{x}),
\label{Survival}
\end{equation}%
where $\alpha (\mathbf{x}):=\exp (\mathbf{x}^{\prime }\boldsymbol{\theta}),$ $%
\boldsymbol{\theta}\in \mathbb{R}^{p}$ is the unknown vector of coefficients and $\mathcal{L}(y;\mathbf{x})$ is some predictor-dependent slowly varying
function, such that $\mathcal{L}(yk;\mathbf{x})/\mathcal{L}(y;\mathbf{x}%
)\rightarrow 1$ for any $k>0$ as $y\rightarrow \infty .$ Specifically, following \citet{Hall(1982)} we characterize the slowly varying function as,
\begin{equation}
\mathcal{L}(y;\mathbf{x}):=c_{0}(\mathbf{x})+c_{1}(\mathbf{x})y^{-\beta ( 
	\mathbf{x})}+o(y^{-\beta (\mathbf{x})})  \label{Slow}
\end{equation}
where $c_{0}(\mathbf{x})$ and $c_{1}(\mathbf{x})$ are functions in $\mathbf{x}$ with $c_{0}(\mathbf{x})>0,$ $\beta(\mathbf{x})>\alpha(\mathbf{x})$ a positive function and $o(y^{-\beta (\mathbf{x})})$ is the higher-order remainder term. As a result, as $y \rightarrow \infty$, $\mathcal{L}(y;\mathbf{x}) \rightarrow c_{0}(\mathbf{x})$ and $\mathcal{\dot{L}}(y;\mathbf{x}) \rightarrow 0$, where $\mathcal{\dot{L}}(y;\mathbf{x}) = \partial \mathcal{L}(y;\mathbf{x})/\partial y.$

From (\ref{Survival}), it follows that the probability density function of $Y_{i}$ conditional on $\mathbf{X}_{i}$ is,%
\begin{equation}
f(y|\mathbf{x};\theta)=\alpha (\mathbf{x})y^{-\alpha (\mathbf{x})-1}\mathcal{%
	L}(y;\mathbf{x})-y^{-\alpha (\mathbf{x})}\mathcal{\dot{L}}(y;\mathbf{x}). \label{fx}
\end{equation}%
Considering (\ref{Slow}) and assuming that $y$ is sufficiently large, it follows that the density in (\ref{fx}) can be approximated as,
\begin{equation}
f(y|\mathbf{x};\theta)\approx c_{0}(\mathbf{x})\alpha (\mathbf{x})y^{-\alpha
	( \mathbf{x})-1};  \label{f_yx}
\end{equation}
see also \citet{WangTsai(2009)}. Thus, the conditional probability function of $Y_i$ given $\mathbf{X}_i$ and $Y_i > y_0$ can be approximated as,
\begin{equation}
f(y|\mathbf{x};\theta)\approx \alpha (\mathbf{x})(y/y_0)^{-\alpha
	( \mathbf{x})-1},  \label{f_y0x}
\end{equation}
where $y_0$ is the threshold that controls the sample fraction used for estimation. Note that (\ref{f_y0x}) is the approximate conditional Pareto density function of an unrestricted random variable and thus, its use when some form of censoring (such as right censoring\footnote{%
	An observation is said to be right censored at $y_c$ if the exact value of the
	observation is not known except that it is greater than or equal to $y_c$.} in the CPS database) is imposed on the data will originate inconsistent tail index parameter estimates. 

In the censored case, rather than observing the outcome $y_{i}$, as in the previous section, we effectively observe $w_{i}$ as defined in (\ref{wi_a}). In this context, the adequately adjusted conditional Pareto density function is,%
\begin{equation}
g(\omega _{i}|\mathbf{x}_{i},y_{c},\boldsymbol{\theta} ):=f(\omega _{i}|\mathbf{x}_{i},y_{c},\boldsymbol{\theta}
)^{I(y_{0}\leq w_{i}<y_{c})}\left[ 1-F(y_{c}|\mathbf{x}_{i},y_{c},\boldsymbol{\theta} )\right]
^{I(w_{i} \geq y_{c})}  \label{log_g}
\end{equation}
where $I(.)$ is the indicator function and $f\left( \left. . \right\vert \mathbf{x}\right) $ and $F\left(
\left. . \right\vert \mathbf{x}\right) $ correspond to the conditional Pareto density function and the conditional cumulative Pareto distribution function, respectively. 

Hence, the negative log-transformed likelihood function for the top-coded data is, 
\begin{equation}
\mathcal{K}_{n}^c(\boldsymbol{\theta };y_c):=\sum_{i=1}^{n}\log g\left( \left. w_{i}\right\vert \mathbf{x}_{i}; y_c;\boldsymbol{%
	\theta }\right) \label{Kn}
\end{equation}
where $g\left( \left. w_{i}\right\vert \mathbf{x}_{i}; y_c;\boldsymbol{\theta }\right)$ is as defined in $(\ref{log_g})$, $w_i$ is given in (\ref{wi_a}) and $y_c$ is the censoring threshold.

Moreover, since
\begin{eqnarray}
log\left[ g(\omega _{i}|\mathbf{x}_{i},y_{c},\boldsymbol\theta )\right]
&=&I(y_{0}\leq w_{i}<y_{c})\log f(\omega _{i}|\mathbf{x}_{i},y_{c},\theta )+I(w_{i}\geq
y_{c})\log \left[ 1-F(y_{c}|\mathbf{x}_{i},y_{c},\theta )\right] \notag\\
&=&I_{\left\{ y_0 \leq w_{i} < y_c\right\} }\{ \log
\alpha \left( \mathbf{x}_{i}\right) +\alpha \left( \mathbf{x}_{i}\right)
\log y_{0}-\left[ \alpha \left( \mathbf{x}_{i}\right) +1\right] \log
w_{i}\} \notag\\
&&+I_{\left\{ w_{i}=y_c\right\} }\{ \alpha \left( \mathbf{x}
_{i}\right) \left[ \log \left( y_{0}\right) -\log (y_c)\right] \},
\label{censored}
\end{eqnarray}%
if we replace $\alpha \left( \mathbf{x}_{i}\right) =\exp \left( \mathbf{x}%
_{i}^{\prime }\boldsymbol{\theta }\right) $ the approximate
negative log-likelihood function in (\ref{Kn}) (omitting for simplicity of notation the terms not related to $\boldsymbol{\theta} $) becomes, 
\begin{eqnarray}
\mathcal{K}_{n}^c(\boldsymbol{\theta} ;y_c) &=&\sum_{i=1}^{n}%
I_{\left\{ y_{0} \leq w_{i}<y_c\right\} }\left( \exp \left( \mathbf{x}%
_{i}^{^{\prime }}\boldsymbol{\theta} \right) \log \left( \frac{w_{i}}{y_{0}}\right) -%
\mathbf{x}_{i}^{^{\prime }}\boldsymbol{\theta} \right)   \notag \\
&&-\sum_{i=1}^{n}I_{\left\{ w_{i}=y_c\right\} }\exp \left( \mathbf{x}%
_{i}^{^{\prime }}\boldsymbol{\theta} \right) \log \left( \frac{y_{0}}{y_c}\right). 
\label{L_cens}
\end{eqnarray}%
Hence, we see from this approximate log-likelihood function that censuring the data imposes a penalty term, which the unrestricted estimator does not take into consideration.

To derive the limit distribution of the parameter estimators and corresponding test statistics we consider, as in \citet{WangTsai(2009)}, the following assumptions:

\bigskip

\noindent \textbf{Assumption A:}

\begin{enumerate}
	\item[(A1)] $n_{0}^{-1}\sum \limits_{i=1}^{n}\mathbf{Z}_{ni}\mathbf{Z}
	_{ni}^{\prime }I\left( w_{i}\geq y_{0}\right)  =\mathbf{\Sigma}_{y_0}^{-1/2}\widehat{\mathbf{\Sigma}}_{y_{0}}\mathbf{\Sigma}_{y_0}^{-1/2}\overset{p}{\rightarrow }\mathbf{I}_{p}$, where $\mathbf{Z}_{ni}:=\mathbf{\Sigma}_{y_0}^{-1/2}\mathbf{x}_i$, $\mathbf{I}_p$ is a $p \times p$ identity matrix and $\mathbf{\widehat{\Sigma}}_{y_{0}}:=n_0^{-1}\sum (\mathbf{x}_i\mathbf{x}'_i)I(w_i \geq y_0).$
	
	\item[(A2)] (\textit{Slowly varying function}) We assume that the remainder term $%
	o(y^{-\beta (\mathbf{x})})$ satisfies \linebreak $\sup_{\mathbf{x}}y^{\beta (\mathbf{x}
		)}o(y^{-\beta (\mathbf{x})})\rightarrow 0$ as $y\rightarrow \infty .$  
	
\end{enumerate}

The following theorem characterizes the limit distribution of the MLE estimates of $\boldsymbol{\theta}$.

\bigskip
\begin{theorem}
	Under Assumptions (A1) - (A2) it follows that 
	\begin{equation*}
	n^{-1/2}\Sigma _{y_{0}}^{-1/2}\Lambda^{-1/2} (\widehat{\boldsymbol{\theta}}-\boldsymbol{\theta} _{0})%
	\overset{d}{\rightarrow }N(\mathbf{0,I}_{p}),
	\end{equation*}%
	where $\Lambda:=E(e_i^2|\mathbf{x}_i)$ and $
	e_{i}=\left\{ 
	\begin{array}{ll}
	\exp \left( \mathbf{x}_{i}^{\prime }\boldsymbol{\theta} \right) \log \left( \frac{%
		w_{i}}{y_{0}}\right) -1 ,\text{ for } & I_{\left\{ y_{0\leq
		}w_{i}<y_c\right\} } \\ 
	-\exp \left( \mathbf{x}_{i}^{\prime }\boldsymbol{\theta} \right) \log \left( \frac{y_{0}}{%
		y_c}\right),\text{ for } & I_{\left\{ w_{i}=y_c\right\} }%
	\end{array}.
	\right.
	$
\end{theorem}

\begin{corollary} Under the same conditions of Theorem 2.1 as $n \rightarrow \infty$ it follows that,
	\begin{equation}
	T_{j}=n^{-1/2}\left(\Sigma _{y_{0},jj}^{-1/2}\right)\Lambda^{-1/2} \widehat{\theta}_{j}\overset{d}{\rightarrow }N(0,1). \label{test}
	\end{equation} 
	where $\Sigma _{y_{0},jj}^{-1/2}$ corresponds to the $(j,j)^{th}$ element of the $\boldsymbol{\Sigma} _{y_{0}}^{-1/2}$ matrix.
\end{corollary}

\subsection{Computation of partial effects}
\label{sec:pe}
A further important and not immediately obvious aspect of the methodology just described relates to the computation of the partial effects of the covariates used in the conditional tail index regression. In specific, for ease of presentation consider%
\begin{equation*}
\overline{F}(y|x; \boldsymbol{\theta}) :=P(Y>y|x; \boldsymbol{\theta}) =\left( \frac{y_0}{y_{i}}\right) ^{\alpha \left( x\right) }.
\end{equation*}%
where for the sake of simplicity but with no loss of generality we consider $x$ to be a scalar and continuous. Thus, to measure the impact of $x$ on $\alpha \left(x\right) $ and subsequently on $\overline{F}(y|x; \boldsymbol{\theta})$, consider
\begin{equation}
\delta :=\frac{\bar{F}\left( \left. y\right\vert \Delta x+x; \boldsymbol{\theta}\right) -\bar{F}
	\left( \left. y\right\vert x; \boldsymbol{\theta}\right) }{\bar{F}\left( \left. y\right\vert
	x; \boldsymbol{\theta}\right) }\times 100 \label{delta}
\end{equation}%
where $y$ is an extreme value, say the ($1-u$) quantile, with $u \in (0,1)$, such that,
\begin{equation}
y=\left( 1-u\right) ^{\frac{1}{\alpha \left( x\right) }}y_{0}.
\end{equation}%
Hence, $\delta$ in (\ref{delta}) measures the probability's percentage variation of an extreme value due to a variation of $x$, $\Delta x$. For example, considering $%
u=0.15$ and $\Delta x=1,$ if $\delta =20\%$ then P$\left(
Y>y_0\right) ,$ where $y_0$ is the 0.85 quantile, increases by 20\% as a result
of $\Delta x=1$. Therefore, the variation of $x$ increases the likelihood of
observing extreme values by 20\%.

For computational purposes, assuming that $\alpha \left( x\right) :=\exp \left( \phi \left( x\right)\right)$, and $\phi(x)$ is some function of $x$, we show in the appendix that
\begin{equation}
\delta \left( u\right) =\left[ \left( 1-u\right) ^{\phi ^{\prime }\left(
	x\right) \Delta x}-1\right] \times 100. \label{du}
\end{equation}%
For instance, in the multivariate case, $\phi \left( 
\mathbf{x}\right) :=\mathbf{x}^{\prime }\mathbf{\beta}$, where $\mathbf{x}$ is a $p\times1$ vector of covariates, $\alpha \left( \mathbf{x}\right) =\exp
\left( \mathbf{x}^{\prime }\mathbf{\beta }\right)$, and the impact of $x_{j}$ on $\alpha(\mathbf{x})$ is,
\begin{equation}
\delta _{j}\left( u\right) =\left[ \left( 1-u\right) ^{\beta _{j}\Delta
	x}-1\right] \times 100. \label{du1}
\end{equation}

Thus, a negative (positive) coefficient increases (decreases) the likelihood of having more extreme values, i.e. $\beta _{j}<0$ ($\beta _{j}>0$) implies $\delta _{j}>0$ ($\delta _{j}<0$) (this is also obvious from the impact on $\alpha(\mathbf{x})$ since if $\alpha(\mathbf{x})$ decreases (increases), the right tail becomes (less)
heavier).

\bigskip

\begin{remark}
	If we use only a portion of the sample to estimate the model, say for example,
	 all observations larger than $y_0,$ then $y$ is the quantile of order $%
	1-u$ of the conditional distribution $P\left( \left. Y<y\right\vert
	Y>y_0\right) ,$ i.e. $P\left( \left. Y<y\right\vert Y>y_0\right) =1-u.$ To determine
	the quantile order of the unconditional distribution $P\left( Y<y\right) $
	we use the relation $P\left( \left. Y<y\right\vert Y>y_0\right)
	=1-u\Rightarrow P\left( Y<y\right) =P\left( Y<y_0\right) +\left( 1-u\right)
	P\left( Y>y_0\right) .$ In the empirical analysis below we use all observations larger than the empirical quantile of 0.80 (see also \cite{Misheletal(2013)}), so that $P\left(
	Y>y_0\right) =0.20$ in the above formula. Hence, when $u=0.15$ and $u=0.20,$ we are
	actually analysing the 96th and 97th quantile, respectively, of the
	unconditional distribution. 
\end{remark}

\bigskip
\begin{remark}
When the covariate considered is discrete (e.g. a dummy variable) a simple adaptation of (\ref{du1}) leads to the following formula, which
	measures the impact of group $d=1$ over $d=0$, $
	\delta \left( u\right) :=\left[ \left( 1-u\right) ^{\frac{\alpha \left( 
			\mathbf{x};d=1\right) }{\alpha \left( \mathbf{x};d=0\right) } -1}-1\right]
	\times 100.$	Given that $\alpha \left( \mathbf{x};d=1\right) $ and $\alpha \left( \mathbf{\ x};d=0\right) $ depend on $\mathbf{x},$ we also need to provide values for $\mathbf{x}$. One possible solution is to replace $\mathbf{x}$ by its
	respective averages.
\end{remark}
\bigskip

\section{Monte Carlo simulation}

In this section we evaluate the finite sample properties of the procedures and their performance in imputing mean wages above the top-code.

\subsection{Finite sample performance of tail index estimators}

To evaluate the performance of the conditional tail index estimator introduced in the previous section, we conduct an in-depth Monte Carlo analysis using several data generation processes (DGPs). In specific, data is simulated from the general framework,%
\begin{eqnarray}
y_{i} &\sim &D\left( \alpha \left( \mathbf{x}_{i}\right) \right) \label{DGP1}\\
\alpha \left( \mathbf{x}_{i}\right) &=&\exp \left( \beta _{1}+\beta
_{2}x_{i}\right) ,\qquad x_{i}\sim U\left( 0,1\right) \label{DGP2}
\end{eqnarray}%
where $\beta _{1}=\beta _{2}=1$ and the $k100$\%, $k\in(0,1)$, largest  observations of the empirical distribution $D(.)$ closely follow a Pareto distribution. We consider the case of right censoring given by the censoring threshold $y_c$ so that the sequence $\left\{ y_{i}\right\} $
is not completely observed. Instead, we observe $ w_{i}=\min \left( y_{i},y_c\right)$. 

To be more precise about the framework used to generate the data, we consider that $D(.)$ in (\ref{DGP1}) is either a Pareto or a Burr distribution\footnote{The cumulative Burr distribution function considered in the simulations is $F(x):=1-(1+x^{-\alpha \rho})^{\frac{1}{\rho}}$ and the corresponding probability density function $f(x):=x^{-1-\alpha \rho}(1+x^{-\alpha \rho})^{-1\frac{1}{\rho}}\alpha$.} and generate samples of size $n\in\{2500, 5000, 10000, 50000\}$. Moreover, we censor the sample considering $y_c=\{\hat{q}_{0.95}^y, \hat{q}_{0.99}^y\}$ which corresponds to the 95th and 99th empirical quantile of $y$. For estimation of the tail index we use the $\lfloor kn\rfloor$ largest observations, with $k=0.2$ when the Pareto distribution is considered and $k=\{0.05, 0.10, 0.20\}$ for the Burr. In the case of samples generated from a Pareto distribution we could have set $k=1$, however, a lower value was considered in order to mimic the conditions typically found in empirical analysis. 

Based on the specifications described above we generated 10,000 sequences of $\left\{ y_{i}\right\} $ and $\left\{w_{i}\right\} $ of size $n$ and use in each iteration three estimation methods:
\begin{itemize}
\item[i)] The tail index regression of \citet{WangTsai(2009)} applied to the sequence of $\left\{ y_{i}\right\}$. We define the resulting estimator as $\hat{\alpha}$. This method should provide the best results since it is applied to the original uncensored data.

\item[ii)] The censored tail index regression introduced in this paper applied to the sequence $\left\{w_{i}\right\} $. The resulting estimator is denoted as $\hat{\alpha}^c$.

\item[iii)] The tail index regression of Wang and Tsai (2009) applied to the censored data $\left\{w_{i}\right\}$. The resulting tail index is defined as $\tilde{\alpha}$. This approach will be useful in providing information on the impact of neglecting the censoring on the tail index estimates.
\end{itemize}

Table 1 provides the bias and RMSEs associated with the estimates of $\beta_1$ and $\beta_2$ in (\ref{DGP2}) computed based on the three approaches described in i), ii) and iii). The first observation we can make is that, in general, the largest bias and RMSEs (regardless of considering $\beta_1$ or $\beta_2$) result from the use of the approach described in iii), i.e., when the censoring is ignored. On the other hand, it is interesting to observe that the difference in the bias and RMSEs obtained from the approaches described in i) and ii) are relatively small, which suggest that the estimation approach which accounts for the censoring produces results close to those obtained when the sample without censoring is used for estimation as is the case in i).

Moreover, this Table also shows that the  bias remains relatively stable and does not
decrease as $n$ increases. There are however different patterns according to
the values of $k$ and $y_{c}.$ For instance, in Cases 3 to 6, which use the Burr
distribution as DGP, a small value of $k$ tends to improve the estimation
results given that the tail of the Burr distribution gets closer to the tail of a Pareto distribution.

To further evaluate the estimation performance of the three estimation approaches in i), ii) and iii), Figure 1 plots the ratios of the RMSEs of the $\alpha$ estimates obtained under the these estimations approaches. In specific, the ratios are,
\[
Ratio\text{ }1=\frac{RMSE\left( \hat{\alpha}^{c}\left( \mathbf{x}_{i}\right)\right) }{RMSE\left( \hat{%
		\alpha}\left( \mathbf{x}_{i}\right)\right) },\qquad Ratio\text{ }2=\frac{RMSE\left( \tilde{\alpha}\left( \mathbf{x}_{i}\right)%
	\right) }{RMSE\left( \hat{\alpha}\left( \mathbf{x}_{i}\right)\right) }.
\]%
Since $\hat{\alpha}$, obtained as described in i), is the best
estimator, $Ratio$ $1$ and $Ratio$ $2$ are larger than 1, across the
different values of $n.$ However, Ratio 1 is just slightly above 1,
which means that the censored estimator performs very well and mimics
closely the behavior of the best estimator, $\hat{\alpha}$, although the former is based on the censored data. On the contrary,
Ratio 2 is substantially higher than 1, which means that, ignoring the censoring
when estimating the tail index produces an inconsistent estimator; see Figure 1.

The censoring threshold $y_{c}$ also impacts the estimation results, i.e., the lower  its value, the greater is the impact of censoring on estimation, and Ratio 2 tends to be larger (see, for example, the results for Case 4 in Table 1 and Figure 1).

\begin{landscape}
	\begin{table}[htbp]
		\caption{Bias and RMSE of estimators}
		\bigskip
		\label{tab:MC2}
\begin{center}
  \scalebox{0.7}{ \begin{tabular}{lSSSSSSSSSSSS}
   		\toprule
    	\multicolumn{13}{c}{\textbf{Case 1: DGP Pareto (k=0.20 and $\mathbf{y_c = Q_y(0.95)}$)}} \\ 
    	 & \multicolumn{1}{c}{$\hat{\alpha}\left( \mathbf{x}_{i}\right)$} &\multicolumn{1}{c}{$\hat{\alpha}^c\left( \mathbf{x}_{i}\right)$}  & \multicolumn{1}{c}{$\tilde{\alpha}\left( \mathbf{x}_{i}\right)$} & \multicolumn{1}{c}{$\hat{\alpha}\left( \mathbf{x}_{i}\right)$} &\multicolumn{1}{c}{$\hat{\alpha}^c\left( \mathbf{x}_{i}\right)$}  & \multicolumn{1}{c}{$\tilde{\alpha}\left( \mathbf{x}_{i}\right)$}& \multicolumn{1}{c}{$\hat{\alpha}\left( \mathbf{x}_{i}\right)$} &\multicolumn{1}{c}{$\hat{\alpha}^c\left( \mathbf{x}_{i}\right)$}  & \multicolumn{1}{c}{$\tilde{\alpha}\left( \mathbf{x}_{i}\right)$}& \multicolumn{1}{c}{$\hat{\alpha}\left( \mathbf{x}_{i}\right)$} &\multicolumn{1}{c}{$\hat{\alpha}^c\left( \mathbf{x}_{i}\right)$}  & \multicolumn{1}{c}{$\tilde{\alpha}\left( \mathbf{x}_{i}\right)$} \\ \midrule
       n   &  \multicolumn{3}{c}{2500} & \multicolumn{3}{c}{5000} & \multicolumn{3}{c}{10000}&\multicolumn{3}{c}{50000} \\
    bias($\beta_1$) & 0.0025 & 0.0004 & 0.4553 & 0.0011 & -0.0002 & 0.4539 & 0.0004 & -0.0005 & 0.4528 & 0.0001 & -0.0002 & 0.4526 \\
   bias($\beta_2$) & 0.0029 & 0.0024 & -0.4241 & 0.0027 & 0.0034 & -0.4237 & 0.0014 & 0.0020 & -0.4246 & 0.0003 & 0.0007 & -0.4257 \\
   RMSE($\beta_1$)& 0.0775 & 0.0937 & 0.4611 & 0.0552 & 0.0660 & 0.4567 & 0.0386 & 0.0465 & 0.4542 & 0.0174 & 0.0208 & 0.4529 \\
  RMSE($\beta_2$)& 0.1703 & 0.1942 & 0.4422 & 0.1208 & 0.1369 & 0.4329 & 0.0852 & 0.0963 & 0.4292 & 0.0378 & 0.0430 & 0.4267 \\
    
          \multicolumn{13}{c}{\textbf{Case 2: DGP Pareto (k=0.20 and $\mathbf{y_c = Q_y(0.99)}$)}} \\ \midrule
       n &  \multicolumn{3}{c}{2500} & \multicolumn{3}{c}{5000} & \multicolumn{3}{c}{10000}&\multicolumn{3}{c}{50000} \\
   bias($\beta_1$) & 0.0025 & -0.0012 & 0.1000 & 0.0011 & -0.0010 & 0.0984 & 0.0004 & -0.0005 & 0.0980 & 0.0001 & -0.0001 & 0.0977 \\
   bias($\beta_2$)& 0.0029 & 0.0070 & -0.1202 & 0.0027 & 0.0050 & -0.1202 & 0.0014 & 0.0023 & -0.1219 & 0.0003 & 0.0005 & -0.1229 \\
  RMSE($\beta_1$) & 0.0775 & 0.0807 & 0.1258 & 0.0552 & 0.0575 & 0.1126 & 0.0386 & 0.0401 & 0.1051 & 0.0174 & 0.0182 & 0.0992 \\
   RMSE($\beta_2$)& 0.1703 & 0.1745 & 0.2006 & 0.1208 & 0.1240 & 0.1659 & 0.0852 & 0.0871 & 0.1460 & 0.0378 & 0.0389 & 0.1281 \\

          \multicolumn{13}{c}{\textbf{Case 3: DGP Burr $\rho$=-2 (k=0.05 and $\mathbf{y_c = Q_y(0.99)}$)}} \\ \midrule

      n  &  \multicolumn{3}{c}{2500} & \multicolumn{3}{c}{5000} & \multicolumn{3}{c}{10000}&\multicolumn{3}{c}{50000} \\
   bias($\beta_1$)  & 0.0015 & -0.0118 & 0.3277 & -0.0006 & -0.0084 & 0.3241 & -0.0025 & -0.0065 & 0.3229 & -0.0042 & -0.0061 & 0.3210 \\
   bias($\beta_2$)& 0.0388 & 0.0488 & -0.3189 & 0.0186 & 0.0242 & -0.3356 & 0.0134 & 0.0161 & -0.3409 & 0.0073 & 0.0097 & -0.3452 \\
   RMSE($\beta_1$)& 0.1450 & 0.1683 & 0.3562 & 0.1021 & 0.1180 & 0.3387 & 0.0719 & 0.0821 & 0.3301 & 0.0323 & 0.0372 & 0.3224 \\
    RMSE($\beta_2$) & 0.4088 & 0.4518 & 0.4549 & 0.2833 & 0.3121 & 0.4043 & 0.1991 & 0.2176 & 0.3754 & 0.0891 & 0.0976 & 0.3523 \\
     
          \multicolumn{13}{c}{\textbf{Case 4: DGP Burr $\rho$=-2 (k=0.10 and $\mathbf{y_c = Q_y(0.95)}$)}} \\ \midrule
     n   &  \multicolumn{3}{c}{2500} & \multicolumn{3}{c}{5000} & \multicolumn{3}{c}{10000}&\multicolumn{3}{c}{50000} \\
    bias($\beta_1$)  & -0.0099 & -0.0239 & 0.9137 & -0.0120 & -0.0258 & 0.9098 & -0.0129 & -0.0259 & 0.9079 & -0.0137 & -0.0259 & 0.9071 \\
    bias($\beta_2$) & 0.0297 & 0.0380 & -0.6485 & 0.0250 & 0.0402 & -0.6489 & 0.0216 & 0.0372 & -0.6512 & 0.0200 & 0.0371 & -0.6519 \\
   RMSE($\beta_1$)& 0.1055 & 0.1596 & 0.9196 & 0.0754 & 0.1129 & 0.9127 & 0.0537 & 0.0820 & 0.9094 & 0.0271 & 0.0434 & 0.9074 \\
    RMSE($\beta_2$)& 0.2609 & 0.3563 & 0.6637 & 0.1835 & 0.2492 & 0.6561 & 0.1305 & 0.1773 & 0.6548 & 0.0602 & 0.0857 & 0.6527 \\

          \multicolumn{13}{c}{\textbf{Case 5: DGP Burr $\rho$=-2 (k=0.20 and $\mathbf{y_c = Q_y(0.99)}$)}} \\ \midrule
       n &  \multicolumn{3}{c}{2500} & \multicolumn{3}{c}{5000} & \multicolumn{3}{c}{10000}&\multicolumn{3}{c}{50000} \\
    bias($\beta_1$) & -0.0364 & -0.0435 & 0.0602 & -0.0378 & -0.0433 & 0.0586 & -0.0385 & -0.0428 & 0.0582 & -0.0386 & -0.0421 & 0.0581 \\
    bias($\beta_2$)& 0.0487 & 0.0578 & -0.0733 & 0.0486 & 0.0558 & -0.0732 & 0.0474 & 0.0531 & -0.0748 & 0.0460 & 0.0510 & -0.0762 \\
   RMSE($\beta_1$)& 0.0847 & 0.0909 & 0.0966 & 0.0665 & 0.0716 & 0.0798 & 0.0542 & 0.0584 & 0.0693 & 0.0422 & 0.0458 & 0.0606 \\
 RMSE($\beta_2$)& 0.1737 & 0.1807 & 0.1739 & 0.1283 & 0.1342 & 0.1344 & 0.0961 & 0.1009 & 0.1089 & 0.0592 & 0.0638 & 0.0840 \\

          \multicolumn{13}{c}{\textbf{Case 6: DGP Burr $\rho$=-2 (k=0.20 and $\mathbf{y_c = Q_y(0.95)}$)}} \\ \midrule
   n &  \multicolumn{3}{c}{2500} & \multicolumn{3}{c}{5000} & \multicolumn{3}{c}{10000}&\multicolumn{3}{c}{50000} \\
   bias($\beta_1$) & -0.0364 & -0.0551 & 0.4134 & -0.0378 & -0.0557 & 0.4119 & -0.0385 & -0.0559 & 0.4108 & -0.0386 & -0.0553 & 0.4109 \\
    bias($\beta_2$) & 0.0487 & 0.0702 & -0.3813 & 0.0486 & 0.0713 & -0.3808 & 0.0474 & 0.0699 & -0.3817 & 0.0460 & 0.0682 & -0.3831 \\
    RMSE($\beta_1$)& 0.0847 & 0.1084 & 0.4196 & 0.0665 & 0.0864 & 0.4150 & 0.0542 & 0.0727 & 0.4124 & 0.0422 & 0.0591 & 0.4112 \\
    RMSE($\beta_2$)& 0.1737 & 0.2040 & 0.4010 & 0.1283 & 0.1532 & 0.3908 & 0.0961 & 0.1182 & 0.3866 & 0.0592 & 0.0804 & 0.3841 \\ \bottomrule
    
    \end{tabular}}
\end{center}
\bigskip
\small{\textbf{Note}: $\hat{\alpha}\left( \mathbf{x}_{i}\right)$ is the tail index regression estimate considering the complete sample of data (with no censoring); $\hat{\alpha}^c\left( \mathbf{x}_{i}\right)$ is the censored tail index regression estimate; and $\tilde{\alpha}\left( \mathbf{x}_{i}\right)$ is the uncensored tail index regression estimate computed from censored data. $n$ corresponds to the total sample size, $k$ to the \% of observations used for the tail index estimation and $\lfloor kn \rfloor$ is the effective number of observations used in the estimation of the tail index. $y_c$ is the censoring value used and $Q_y(\tau)$ corresponds to the $\tau^{th}$ quantile of $y$.}
\end{table}
\end{landscape}

\FloatBarrier

\begin{figure}[h!]
	\caption{Ratios of the tail index estimates' RMSEs}
	\label{f1}
	\begin{center}
		$%
		\begin{array}{cc}
		\epsfxsize=3.1in \epsffile{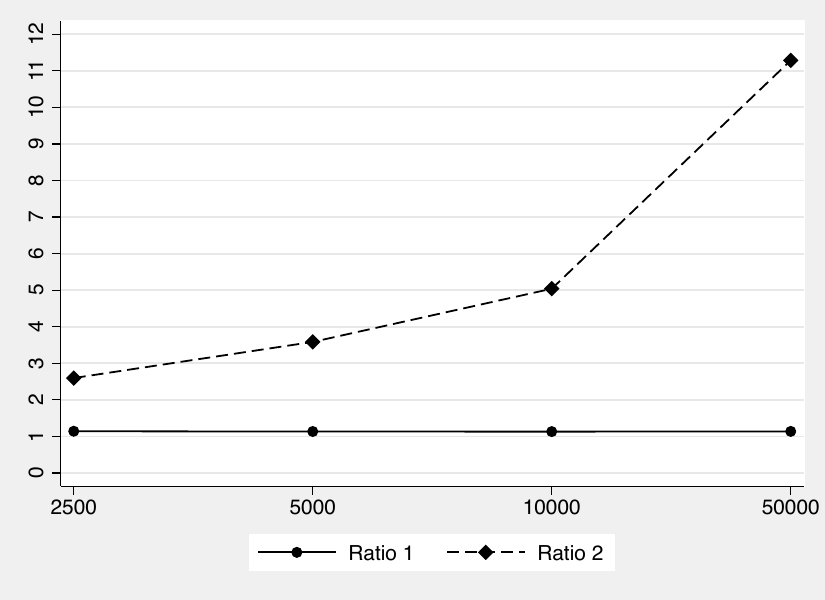} &  \epsfxsize=3.1in \epsffile{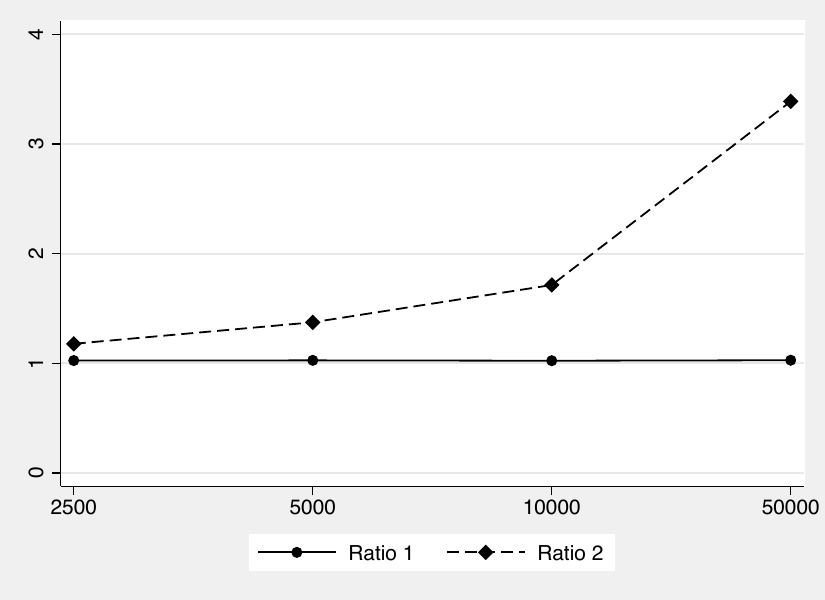}  \\
		\multicolumn{1}{l}{\mbox{\small DGP1: Pareto (k=0.20 and $y_c = Q_y(0.95)$)}} &  \multicolumn{1}{l}{\mbox{\small DGP2: Pareto (k=0.20 and $y_c = Q_y(0.99)$)}} \\%
		\epsfxsize=3.1in \epsffile{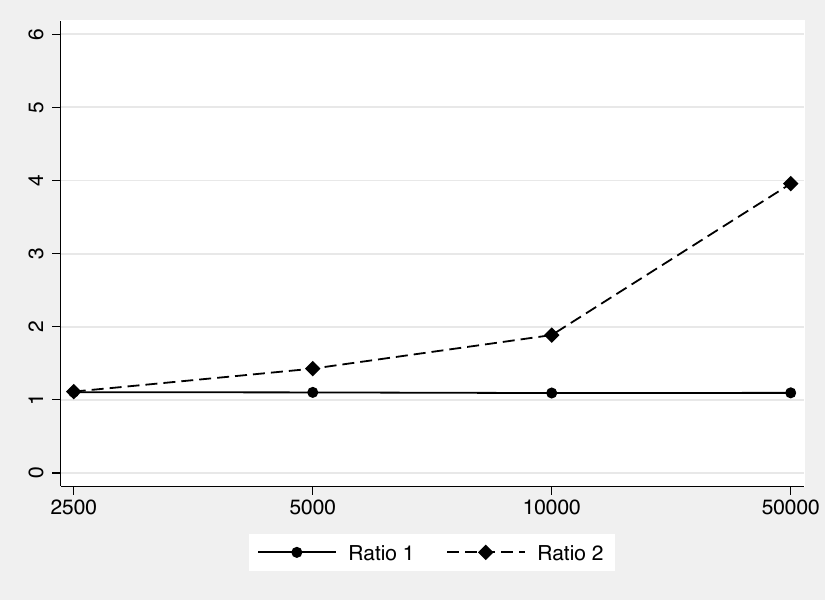} &  \epsfxsize=3.1in \epsffile{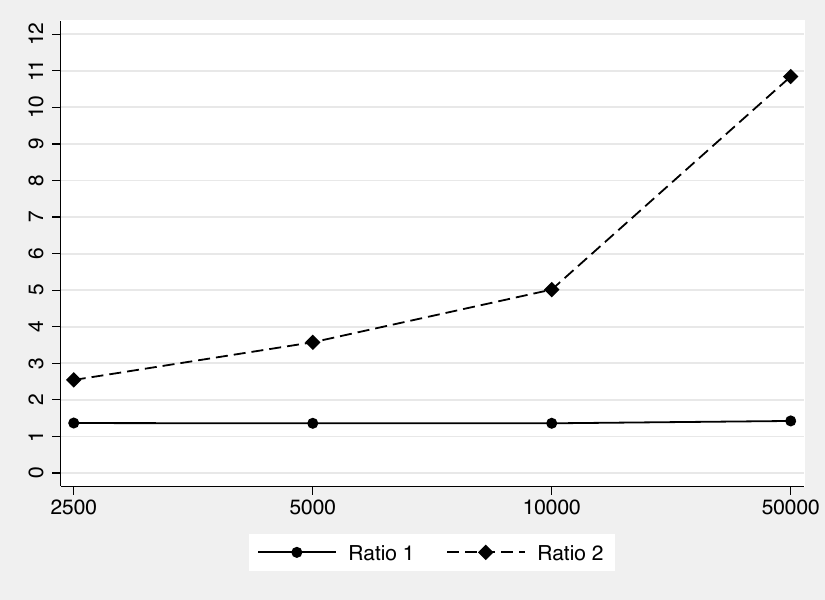}  \\
		\multicolumn{1}{l}{\mbox{\small DGP3: Burr $\rho$=-2 (k=0.05 and $y_c = Q_y(0.99)$)}} &  \multicolumn{1}{l}{\mbox{\small DGP4: Burr $\rho$=-2 (k=0.10 and $y_c = Q_y(0.95)$)}} \\%
		\epsfxsize=3.1in \epsffile{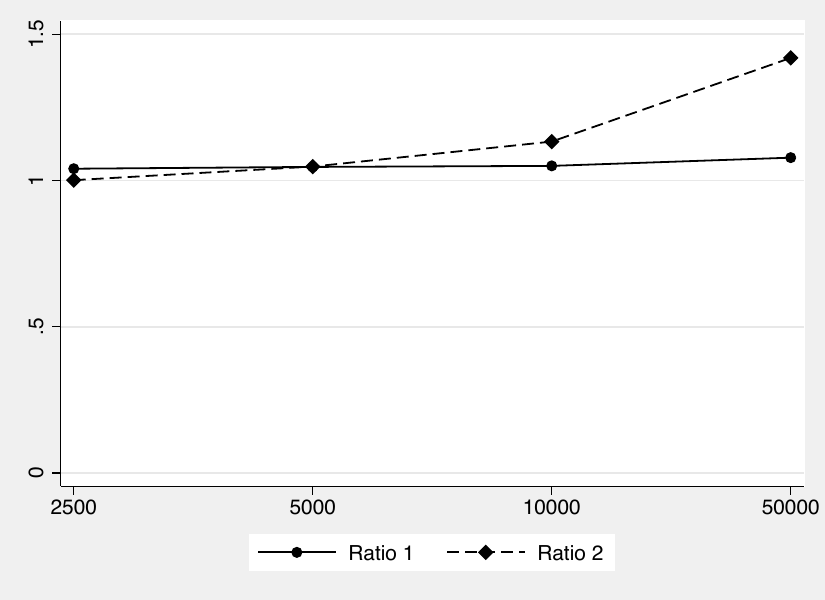} &  \epsfxsize=3.1in \epsffile{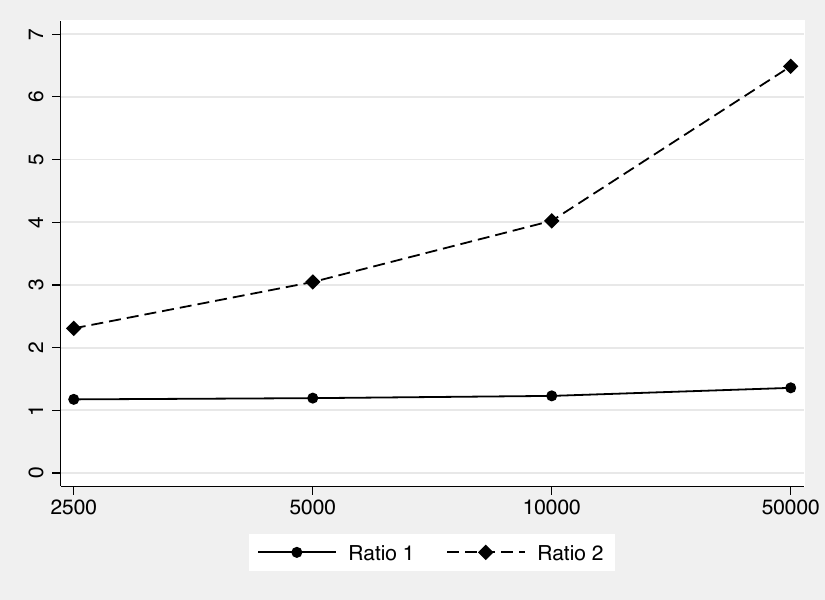}  \\
		\multicolumn{1}{l}{\mbox{\small DGP5: Burr $\rho$=-2 (k=0.20 and $y_c = Q_y(0.99)$)}} &  \multicolumn{1}{l}{\mbox{\small DGP6: Burr $\rho$=-2 (k=0.20 and $y_c = Q_y(0.95)$)}} \\%
		\end{array}%
		$
	\end{center}
	\begin{minipage}{16.5cm}
		\begin{spacing}{0.8}
		\small{\textbf{Note:} $Ratio\text{ }1={RMSE\left( \hat{\alpha}^{c}\right) }/{RMSE\left( \hat{%
					\alpha}\right) }$ and $Ratio\text{ }2={RMSE\left( \tilde{\alpha}%
				\right) }/{RMSE\left( \hat{\alpha}\right) }$.  Ratio 1 compares the censored estimator with the best estimator $\hat{\alpha}$, while Ratio 2 compares the estimator that ignores the censoring with the best estimator $\hat{\alpha}$. We report the results using the Pareto and the Burr distributions as DGP.}
		\end{spacing}	
	\end{minipage}
\end{figure}

\FloatBarrier
\newpage

\subsection{Imputing mean wages}
To provide further insights on the usefulness of the procedure introduced in this paper we provide next an analysis of the performance of the different methods for imputing mean wages. In specific, we compare the following three methods:

\begin{itemize}
	\item[i)] the Pareto-imputed mean wage above the top-code $y_c$,%
\begin{equation}
	\hat{\tau}_{1}\left( y_{c}\right) =\frac{\hat{\alpha}_{1}}{\hat{\alpha}_{1}-1%
	}y_{c} \label{tau1}
\end{equation}
	where $\hat{\alpha}_{1}$ is the tail index estimate considering an uncensored Pareto distribution as in Section 2 (see e.g. \cite{Hill(1975)} and \citet{NicolauRodrigues(2019)} for tail index estimators) and $y_{c}$ is the top-code threshold; 
	
	\item[ii)] the imputed mean wage based on the approach suggested by \cite{Armouretal(2016)}
	\begin{equation}
	\hat{\tau}_{2}\left( y_{c}\right) =\frac{\hat{\alpha}_{Hill}^c}{\hat{\alpha}_{Hill}^c-1%
	}y_{c} \label{tau2}
	\end{equation}
	where $\hat{\alpha}_{Hill}^c$ is a consistent estimate of the tail index
	parameter computed as in (\ref{5}). Note that $\tau_{2}\left( y_{c}\right) :=E\left(y_i|y_i>y_c\right)$ ;
	
	\item[iii)] the imputed mean wage based on the method introduced in this paper%
	\begin{equation}
	\hat{\tau}_{3}\left( y_{c}\right) =\frac{\hat{\alpha}\left( \mathbf{x}%
		_{i}\right) }{\hat{\alpha}\left( \mathbf{x}_{i}\right) -1}y_{c} \label{tau3}
	\end{equation}
	where $\hat{\alpha}\left( \mathbf{x}_{i}\right) =\exp \left( \mathbf{x}%
	_{i}^{\prime }\boldsymbol{\hat{\theta}}\right) .$ Note that $\tau _{3}\left(
	y_{c}\right) :=E\left( \left. y_{i}\right\vert \mathbf{x}%
	_{i}, y_i>y_c\right) .$
\end{itemize}

The main difference between $	\hat{\tau}_{3}\left( y_{c}\right)$ and the other two approaches ($	\hat{\tau}_{1}\left( y_{c}\right)$ and 	$\hat{\tau}_{2}\left( y_{c}\right)$) is that in the former the particular characteristics of the
individuals whose wage is above the
threshold are taken into account through $\mathbf{x}_{i}$. Interestingly, the estimator $\hat{\tau}_{3}\left( y_{c}\right)$ corresponds to the optimal mean
square predictor because $\tau_{3}\left( y_{c}\right)$ is the conditional expectation of $y_{i}$
given $\mathbf{x}_{i}$ and $y_i>y_c$. This follows from the well known result $E \left( y_{i}-E\left( \left. y_{i}\right\vert \mathbf{x}_{i},y_i>y_c\right) \right) ^{2} \leq E \left( y_{i}-g\left(
. \right) \right) ^{2}$ where $g\left( . \right) $ is
any other predictor of $y_{i}$ given $y_i>y_c$. It turns out that $\tau_{2}\left( y_{c}\right)$ is optimal
only if $y_{i}$ is mean-independent of $\mathbf{x}_{i},$ in which case
both $\tau_{2}\left( y_{c}\right)$ and $\tau_{3}\left( y_{c}\right)$ coincide. 

Thus, having established the superiority of $\tau_{3}\left( y_{c}\right)$, it remains
to be shown how much improvement is provided by $\tau_{3}\left( y_{c}\right)$ compared to $\tau_{1}\left( y_{c}\right)$ and $\tau_{2}\left( y_{c}\right)$ when computing the mean wage above $y_c$. The following Monte
Carlo study tries to answer this question.  Our experiment is based on the
following steps:

\newpage

\begin{enumerate}
	\item Select a sample size from
	\[
	n \in \left\{ 250,500,1000,2000,5000,20000\right\};
	\]
	
	\item Simulate $y_{i},$ $i=1,2,...,n$ according to a conditional Pareto distribution $%
	P\left( \alpha \left( \mathbf{x}_{i}\right) \right) $ where $\alpha \left( 
	\mathbf{x}_{i}\right) =\exp \left( 1+2x_{i}\right) $ and $x_{i}\sim U\left(
	0,1\right) .$ For each $i=1,2,...,n$ simulate $\alpha \left( \mathbf{x}_{i}\right) $
	and then the corresponding $y_{i}$;
	
	\item All observations above quantile $0.95$, which corresponds to $y_{c},$ are
	censored, but their original values are saved for comparison purposes (these values are used to 
	assess the estimators' predictive precision). The  data used for estimation are $\left\{
	w_{i},i=1,2,...,n\right\} $ where $w_{i}=\min \left( y_{i},y_{c}\right) ,$
	from which we estimate $\hat{\tau}_{1}\left( y_{c}\right) ,$ $\hat{\tau}%
	_{2}\left( y_{c}\right) $ and $\hat{\tau}_{3}\left( y_{c}\right) .$
	
	\item The estimators $\hat{\tau}_{1}\left( y_{c}\right) ,$ $\hat{\tau}%
	_{2}\left( y_{c}\right) $ and $\hat{\tau}_{3}\left( y_{c}\right) $ are used
	to predict the imputed mean value above the threshold $y_{c}.$ 
	
	\item Steps 2 to 4 are repeated 2000 times and the mean square errors (MSE) of $\hat{\tau%
	}_{1}\left( y_{c}\right) ,$ $\hat{\tau}_{2}\left( y_{c}\right) $ and $\hat{%
		\tau}_{3}\left( y_{c}\right) $ are computed.
\end{enumerate}

Other combinations of $\alpha \left( \mathbf{x}_{i}\right) $ produce
essentially the same results as long as $\alpha \left( \mathbf{x}_{i}\right)
\geq 1$, and for this reason we present results only for the case $\alpha \left( \mathbf{x%
}_{i}\right) =\exp \left( 1+2x_{i}\right) $ (note that $E\left(
\alpha \left( \mathbf{x}_{i}\right) \right) =3.7).$ However,  $\alpha \left( \mathbf{x}_{i}\right) $ should be set so that $\alpha\left( \mathbf{x}_{i}\right) >1,$ otherwise the conditional and marginal
expected values do not exist, and consequently  none of the above estimators will be
well defined. The case $0<\alpha \left( \mathbf{x}_{i}\right) <1$ should
be dealt with using other estimators, such as, for example, the median, $\sqrt[{\hat{\alpha}\left(\mathbf{x}_{i}\right)}]{2}y_{c}$). 

Figure 2 illustrates our results. The lines represent two MSE ratios,  $Ratio\text{ } 1:=MSE\left( \hat{\tau}_{1}\right) /MSE\left( \hat{\tau}_{3}\right)$ and $Ratio\text{ }2:=MSE\left( \hat{\tau}_{2}\right) /MSE\left( \hat{\tau}_{3}\right)$, computed over different sample sizes (n = (250, 500, 1000, 2000, 5000, 10000)).

\begin{figure}[htp]
\begin{center}
		\caption{\label{Fig2}MSE Ratios Computed Over Different Sample Sizes}
\includegraphics[height=3in, width=0.65 \textwidth]{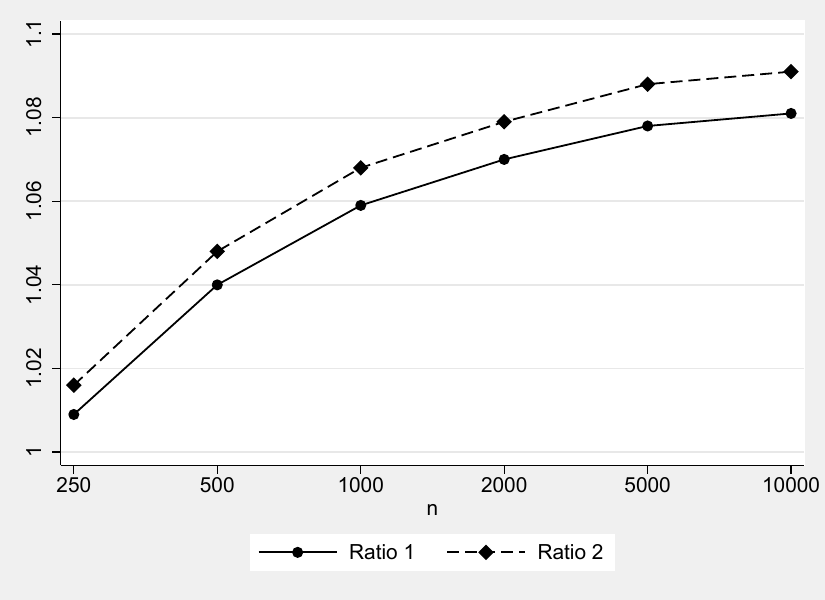}
\end{center}
	\end{figure}

\FloatBarrier

\newpage

As can be observed from Figure \ref{Fig2}, the $\hat{\tau}_{3}(y_c)$ estimator introduced in this paper produces the best
results as all MSE ratios are larger than one. The gains are modest
(between 1\% and 2\%) when the sample size is small (n=250), but they increase
steadily as the sample size increases. Another conclusion, is that $\hat{\tau}_{2}$ (\cite{Armouretal(2016)}) is
better than the (naive) $\hat{\tau}_{1}$ estimator that does not accommodate the censoring.

\section{Empirical analysis}

In this section, we use censored publicly available CPS data to evaluate how the right tail index of the US wage distribution has changed over time and how these changes may differ across the characteristics of individuals, occupations and industries.  In specific, we show that the new tail index estimator introduced provides very rich and detailed insights about the right tail distribution of wages. We also assess the sensitivity of the adjustment of the top coded wage to changes over time and across the characteristics of individuals. 

\subsection{Data}

For the empirical analysis the March CPS files from IPUMS for the period between 1992 and 2017 are used. The wage measure is  top-coded at $\$$1923 between 1989 and 1997, and at $\$$2884 between 1998 and 2017. The sample is restricted to workers between 16 to 64 year-old on full-time full year basis employed during the CPS sample survey reference week (35+ hours per week, 40+ weeks per year). 
Following \citet{AutorDorn(2013)} the real weekly wage data are weighted by the appropriate CPS weight to provide a measure of the full distribution of weekly wages paid.\footnote{Wages are converted to 2012 values using the GDP personal consumption expenditure deflator.}$^,$\footnote{Workers in our sample come from outgoing rotation groups 4 and 8 and according to Unicon: When the Outgoing Rotation files are produced, two rotations are extracted from each of the twelve months and gathered into a single annual file. The weights on the file must be modified by the user before giving reliable counts. Since the final weight is gathered from 12 months but only 2/8 rotations, the weight on the outgoing file should be divided by 3 (12/4) before it is applied. The earner weight is gathered from 12 months from the 2 rotations. Since those two rotations were originally weighted to give a full sample, the earner weight must be divided by 12, not 3.} 

Occupations are defined as job task requirements of the US Department of Labor Dictionary of Occupational Titles (DOT, 1977) and Census occupation classifications for routine, abstract and manual task classifications (\citet{AutorDorn(2013)}).

\begin{figure}[h!]
	\caption{Annual unconditional histograms of the weekly wage.}
	\label{f3}
	\begin{center}
		$%
		\begin{array}{cc}
		\epsfxsize=3.1in \epsffile{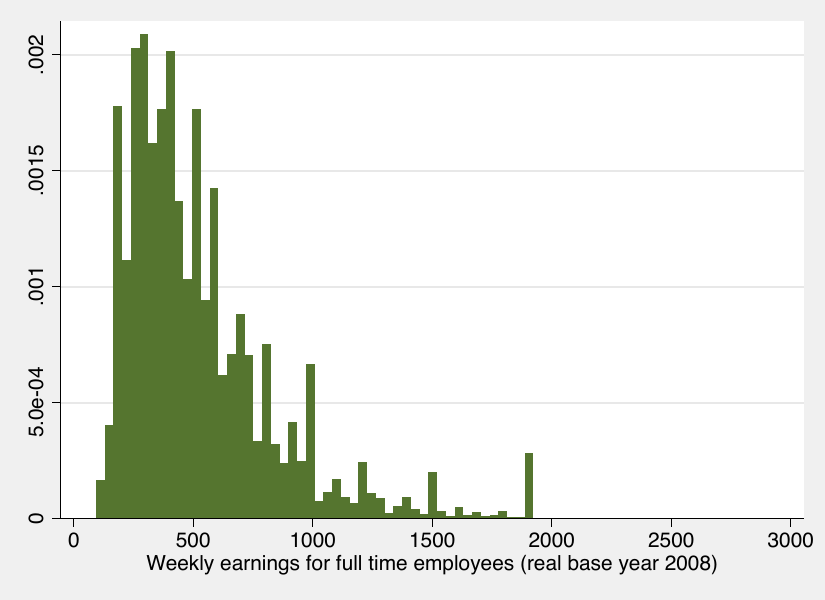} &  \epsfxsize=3.1in \epsffile{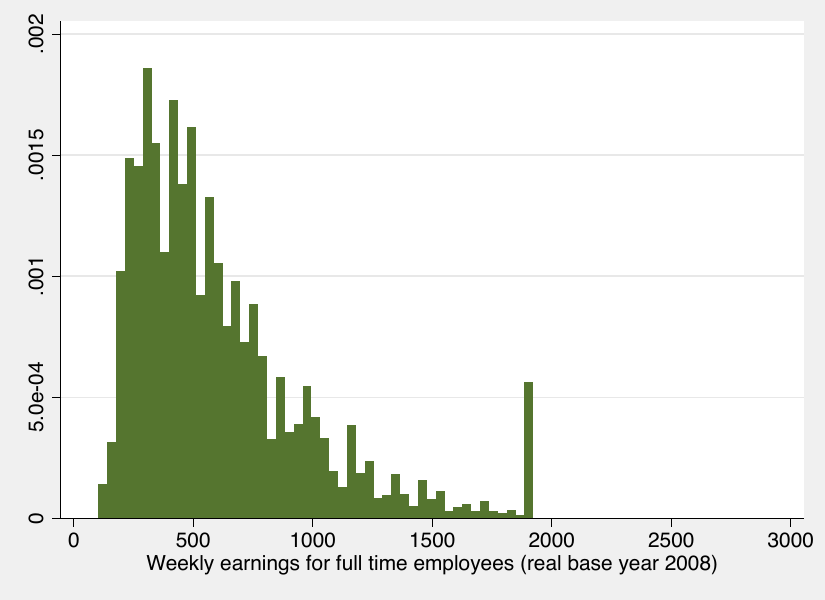}  \\
		\multicolumn{1}{c}{\mbox{\bf (a) 1992}} &  \multicolumn{1}{c}{\mbox{\bf (b) 1997}} \\%
		\epsfxsize=3.1in \epsffile{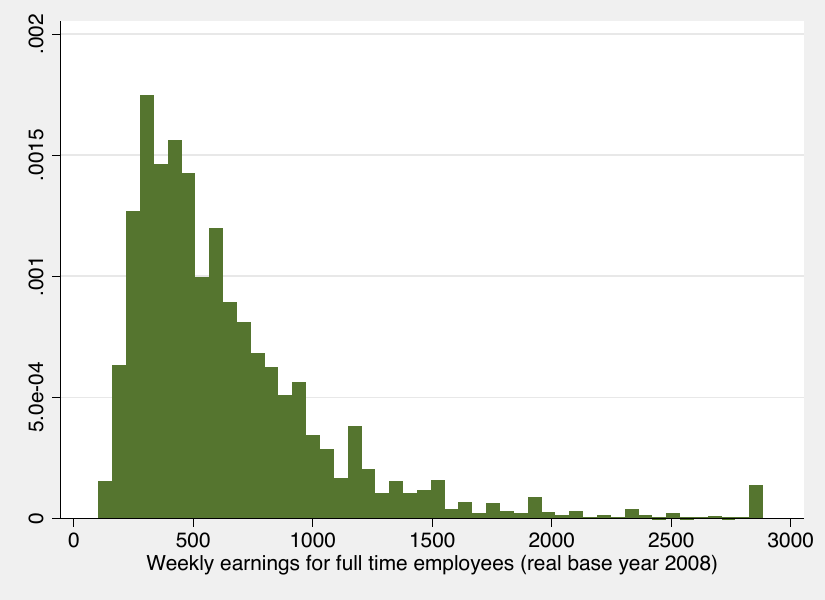} &  \epsfxsize=3.1in \epsffile{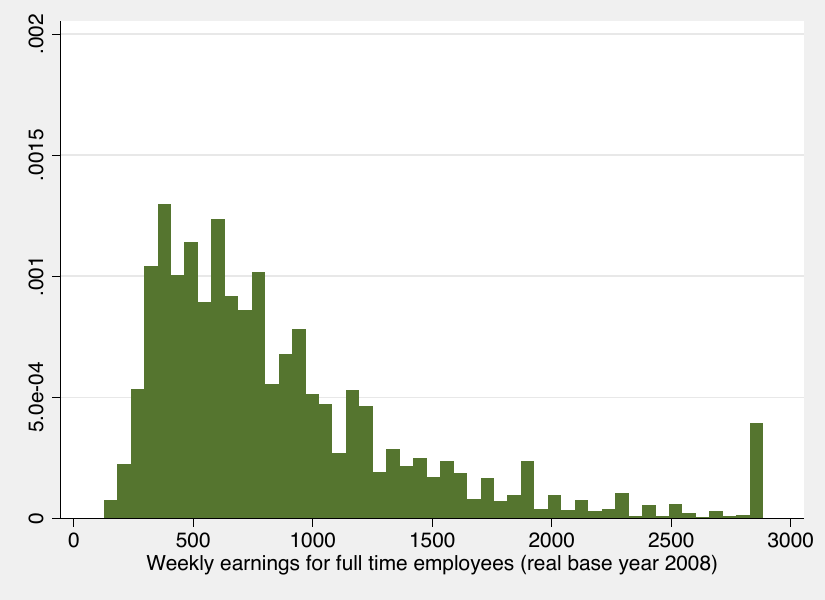}  \\
		\multicolumn{1}{c}{\mbox{\bf (a) 1998}} &  \multicolumn{1}{c}{\mbox{\bf (b) 2007}} \\%
		\epsfxsize=3.1in \epsffile{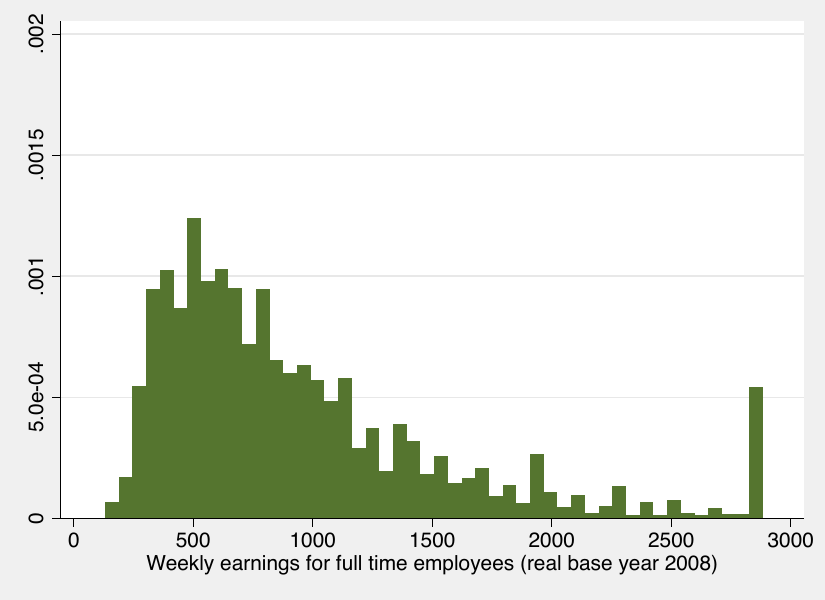} &  \epsfxsize=3.1in \epsffile{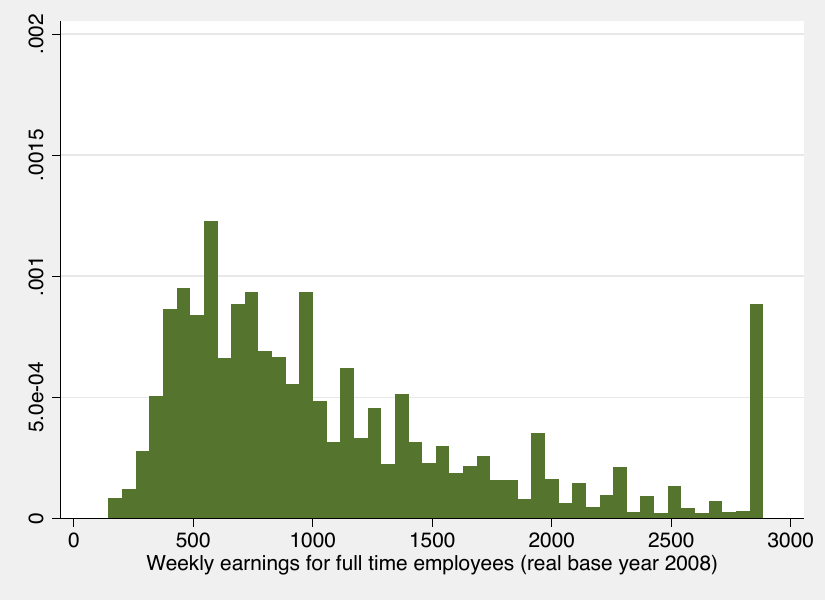}  \\
		\multicolumn{1}{c}{\mbox{\bf (c) 2010}} &  \multicolumn{1}{c}{\mbox{\bf (d) 2017}} \\%
		\end{array}%
		$
	\end{center}
\end{figure}

In Figure \ref{f3} we present the distribution of the weekly wages for 1992, 1997, 1998, 2007, 2010  and 2017. From 1992 to 2017 the concentration of wages has become more skewed to the right. Between 1992 and 1997 the mass points around the top-code increased and with the relaxing of the top-code in 1998 real values beyond that top-code are potentially observable. The same phenomenon also occurs in the most recent period. In 2017 the mass point around the current top-code used in the CPS data is much larger.   

\FloatBarrier

The means and proportions of workers according to their characteristics, occupations and industry, for observations above the 80th percentile, are presented in Table \ref{t2}. In contrast to 1992, in 2017 the population in the right tail is older (41.03 years on average in 1992
and 43.83 years in 2017), the percentage of women is larger (25\% in 1992 increased to 33\% in
2017), and is about one year more educated (15.20 years of education in 1992 and 16.06 years in 2017).

\begin{table}[h!]
	\begin{center}							
		\caption{\textbf{Characteristics of individuals: means and proportions}}\label{t2}	
		\begin{tabular}{llrr}
			\hline
			& & \multicolumn{2}{c}{Year} \\
			Mean & & 1992&2017 \\
			\hline
			\multicolumn{2}{l}{Age } &41.03&43.83\\
			\multicolumn{2}{l}{Female} &0.25&0.33 \\
			\multicolumn{2}{l}{Education} &15.20 &16.06 \\
			\multicolumn{2}{l}{\textbf{Race}}&&\\
			&White &0.90&0.83\\
			&Black &0.05&0.05\\
			& Other race &0.05&0.12\\
			\multicolumn{2}{l}{\textbf{Marital Status}} & &\\
			&Married &0.83&0.81\\
			&Married no spouse&0.01&0.02\\
			&Separated &0.02&0.01\\
			&Divorced&0.09&0.09\\
			&Widowed&0.01&0.01\\
			&Single&0.14&0.16 \\
			\multicolumn{2}{l}{\textbf{Occupations}}&&\\
			&Managers&0.72&0.81\\
			&Administrative&0.10&0.08\\
			&Low skill &0.01&0.01\\
			&Craft &0.04&0.01\\
			&Operators&0.02&0.01\\
			&Transportation&0.11&0.08 \\
			\multicolumn{2}{l}{\textbf{Industry$^a$}}&&\\
			&Agriculture&0.02&0.03\\
			&Construction&0.05&0.05\\
			&Manufacturing&0.22&0.14\\
			&Transports&0.11&0.08\\
			&Trade&0.11&0.09\\
			&Finance&0.09&0.10\\
			&Repair&0.04&0.10\\
			&Personal&0.28&0.32\\
			&Public&0.08&0.09 \\
			\multicolumn{2}{l}{Observations} & 112,960& 64,002 \\
			\multicolumn{2}{l}{Observations above Percentile 80} & 22,485& 12,791 \\
			\hline
		\end{tabular}
		\begin{minipage}{15cm}
			\medskip
			$^a$ \footnotesize{In Appendix C we provide a detailed description of the industry classification.}
			\begin{spacing}{0.8}
				\bigskip
				\small{\textbf{Notes}: This Table reports the means for the variables used in the analysis for both 1992 and 2017. These statistics were calculated for the right tail (for observations above percentile 80). All variables are reported on a scale between 0 and 1 with the exception of age and education which are reported in years. The occupation dummies using 6 aggregate occupation groups are based on the International Standard Classification of occupations (ISCO) as used in \cite{AutorDorn(2013)}. The category Managers includes management, professional, technical, financial sales and public security occupations. The category Administrative consists of routine non cognitive occupations and includes administrative support and retail sales occupations. The category Low-skill includes low-skill services, such as cleaning, guard, food, health, janitors, beauty, recreation, working with children and other personal low-skill services. The category Craft aggregates precision production and craft occupations. The category Operators refers to machine operators, assemblers and inspectors. Finally, the category Transportation includes transportation, construction, mechanics, mining and agricultural occupations.}
			\end{spacing}	
		\end{minipage}
		
	\end{center}	
\end{table}			

\FloatBarrier

The results in Table \ref{t2} show that the 7\% decrease of white individuals in the right tail in 2017 when compared to 1992 seems to be compensated by a 7\% increase of individuals from other races (non-white nor non-black). There is a 3\% change in the composition of the sample, where the reduction of married individuals is compensated by an increase in individuals that are single. However, married still represent the marital status of the majority of individuals in the right tail (83\% in 1992 and 81\% in 2017). 

A further observation that can be made from the results in Table \ref{t2} is job polarization. A significant growth in employment in the right tail
for non-routine cognitive tasks is observed in detriment of  routine occupations (see also \cite{Autor(2019)} and \cite{GoosManning(2007)}). The decrease in the share of employment is even more significant for
non-routine manual tasks (individuals in occupations associated with
transportation, construction and mechanics (Transportation) decreased their share of employment in the right tail, from 11\% in 1992 to 8\% in 2017). The percentage of individuals in occupation Managers, which is the occupation with the largest proportion of individuals, increased from 72\% in 1992 to 81\% in 2017. 

In 1992, the proportion of individuals, working in manufacturing (transports) was 22\% (12\%) and this proportion decreased to 14\% (8\%). This decrease in the share of employment was compensated by an increase in the repair (+6\% between 1992 and 2017) and finance, personal and public industries (+6\% between 1992 and 2017). Note that the industries with the largest number of individuals in the right tail in (1992, 2017) are Personal (28\%, 32\%), Manufacturing (22\%, 14\%), Finance (9\%, 10\%), Repair (4\%, 10\%) and Public (8\%, 9\%). However, Manufacturing (22\%, 14\%), Transport (11\%, 8\%) and Trade (11\%, 9\%) see their weight decrease in 2017.

To illustrate the evolution of the proportions of individuals in the different percentiles  of the overall wage distribution, Figure 4 plots the proportions in percentiles 0.05 to 0.95 considering different attributes, occupations and industries (in the appendix we present additional plots for all other cases analysed in Table \ref{t2}). From this Figure we distinguish two patterns from 1992 to 2017: Other Race, Female, Single and Personal increase in proportion from 1992 to 2017 across all percentiles, whereas Administration and Trade decrease across all percentiles. The number of Black individuals seems to decrease up to around percentile 80 and increases thereafter.

Moreover, this Figure also shows that individuals that are Black, Female or Single as well as individuals working in Administration and Trade display a downward trend over the percentiles, whereas the number of individuals of Other Races and individuals working  in the Personal industry display a different pattern. The former is relatively constant across all percentiles in 1992 but increases for percentiles larger than the median in 2017, and the latter is relatively constant across all percentiles in 1992 and 2017.  Interestingly, Finance shows a different pattern when compared to all other covariates. In specific, the largest proportions are observed at the higher percentiles (i.e. from the median onward). This pattern is very similar across all years analysed.

\begin{figure}[h!]
	\caption{Proportion of individuals in different wage percentiles}
	\label{f4}
	\begin{center}
		
		\scalebox{0.85}{		$
			\begin{array}{cc}
			\epsfxsize=3.1in \epsffile{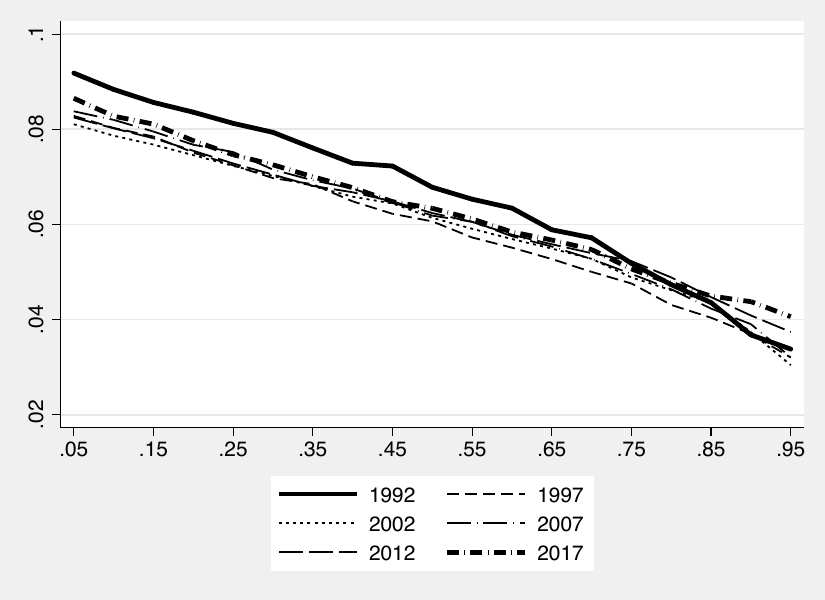}&  \epsfxsize=3.1in \epsffile{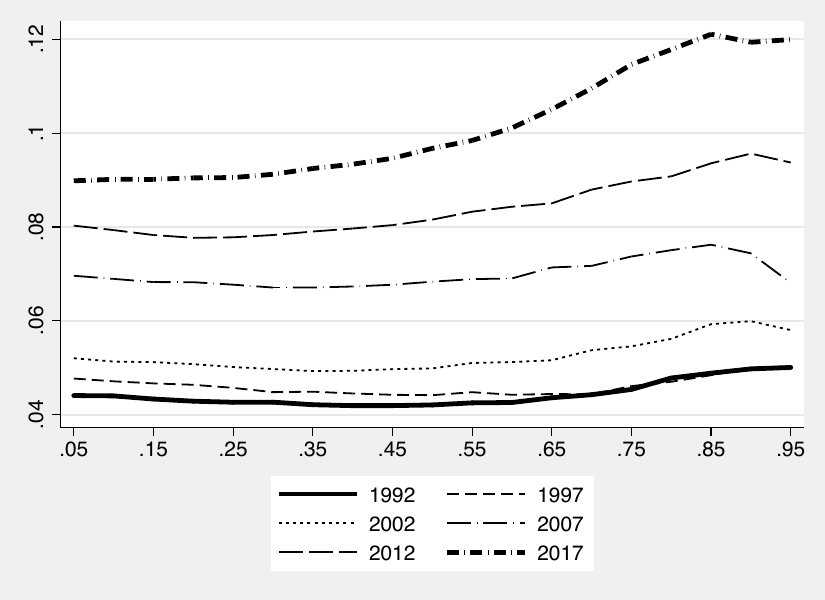}  \\
			\multicolumn{1}{c}{\mbox{\bf Black} }&  \multicolumn{1}{c}{\mbox{\bf Other Race}} \\%
			\epsfxsize=3.1in \epsffile{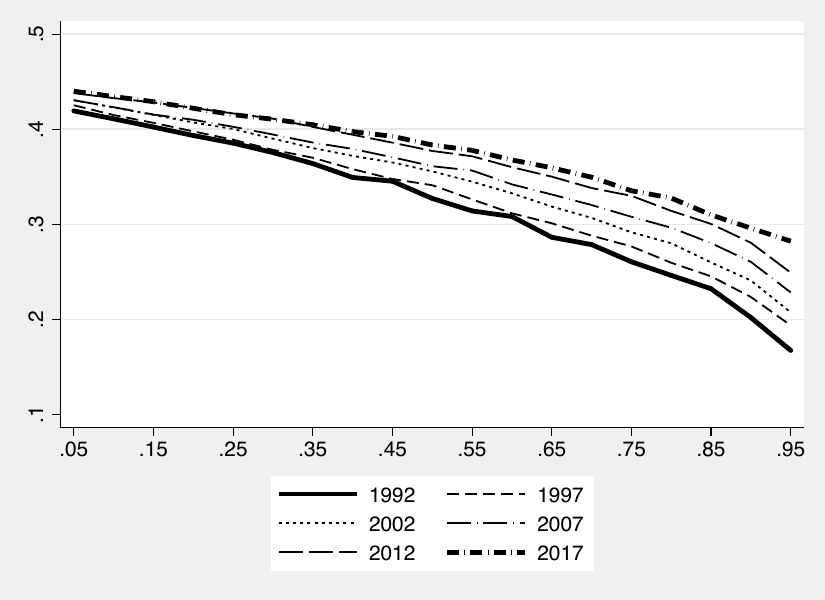} &\epsfxsize=3.1in \epsffile{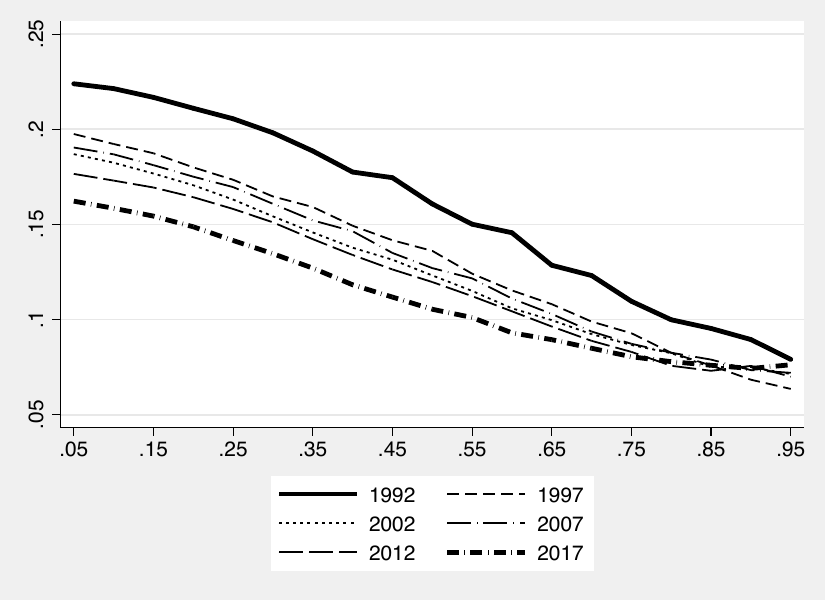}   \\
			\multicolumn{1}{c}{\mbox{\bf Female}} &  \multicolumn{1}{c}{\mbox{\bf Occ: Administration}}   \\%
			\epsfxsize=3.1in \epsffile{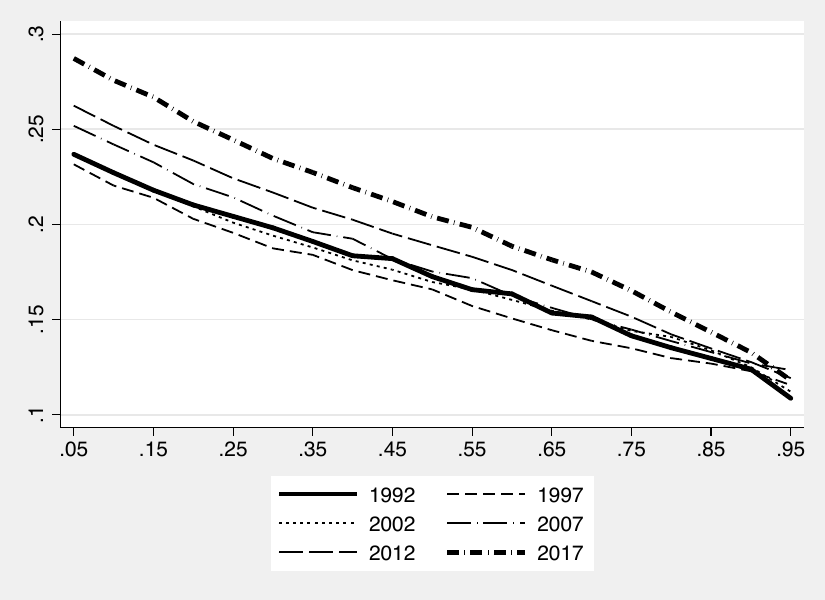} &  \epsfxsize=3.1in \epsffile{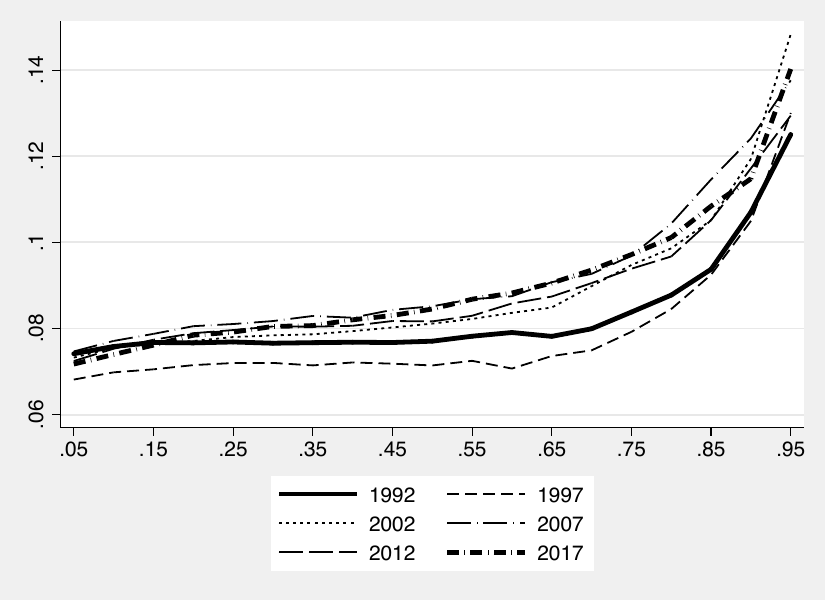}  \\
			\multicolumn{1}{c}{\mbox{\bf Single}} & \multicolumn{1}{c}{\mbox{\bf Ind: Finance}}   \\%
			\epsfxsize=3.1in \epsffile{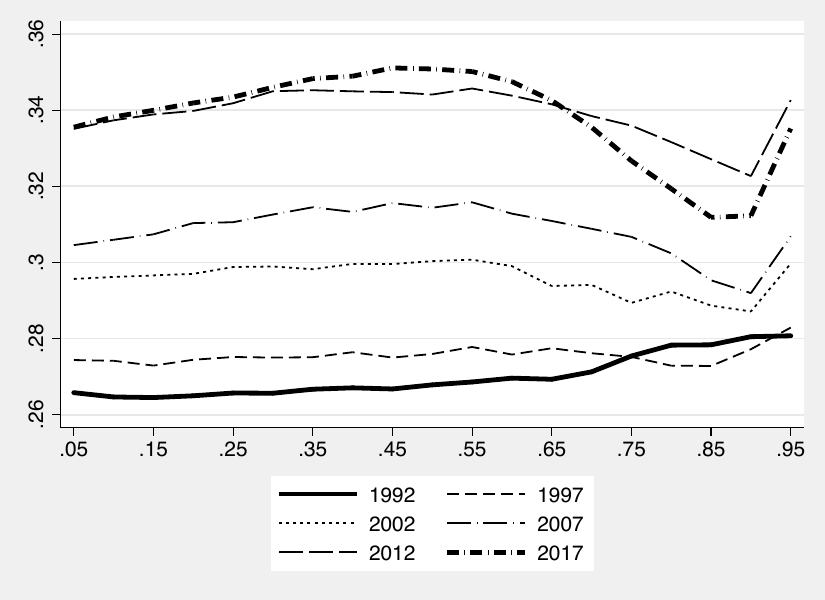}  &  \epsfxsize=3.1in \epsffile{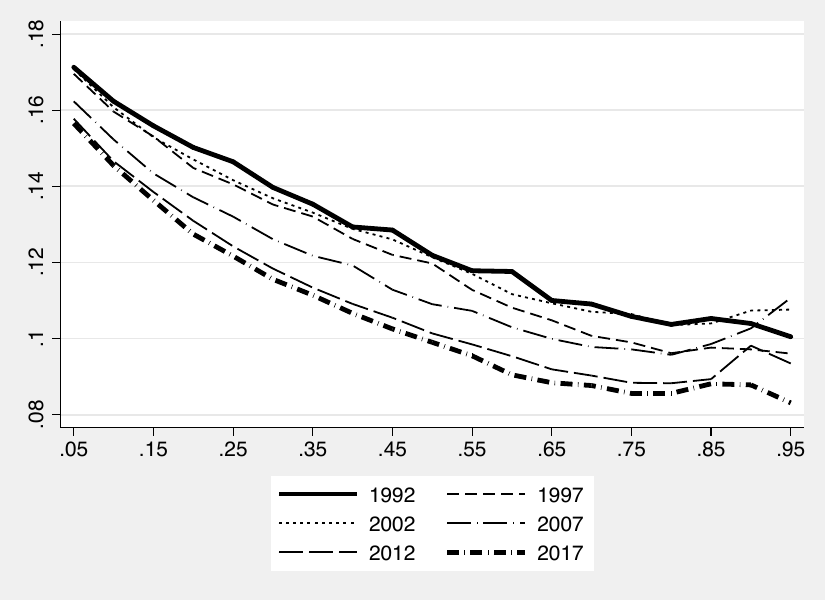}  \\
			\multicolumn{1}{c}{\mbox{\bf Ind: Personal}} &  \multicolumn{1}{c}{\mbox{\bf Ind: Trade}} \\%
			\end{array}%
			$}
	\end{center}

	\begin{minipage}{16.5cm}
		\begin{spacing}{1.25}
			\bigskip
			\small{\textbf{Notes}: The graphs presented in this figure represent the proportion of individuals in different categories for percentiles 0.05, 0.1, 0.15,..., 0.95, computed for the years of 1992, 1997, 2002, 2007, 2012 and 2017.}
		\end{spacing}	
	\end{minipage}
\end{figure}

\FloatBarrier

\subsection{Conditional tail index estimation results}

Table 3 presents the censored and uncensored tail index regression results for 1992 and 2017. A negative (positive) regression coefficient corresponds to a decrease (increase) in the tail index (ceteris paribus) and hence a larger (smaller) number of extreme values may result as a consequence of changes in the specific variable associated to such a coefficient. Before analyzing the partial effects, as described in Section 2.2, the behavior of a tail index estimate computed as an average of the conditional tail indexes is examined, i.e.,
\begin{equation}
\hat{\alpha}_{t}=\frac{1}{n}\sum_{i=1}^{n}\hat{\alpha}\left( \mathbf{x}%
_{t,i}\right) ,\text{ }t=1992,...,2017 \label{alpha_t}
\end{equation}
where $\hat{\alpha}\left( \mathbf{x}_{t,i}\right) =\exp \left( \mathbf{x}%
_{t,i}^{\prime }\mathbf{\hat{\theta}}_{t}\right) $ (we only report results for 1992 and 2017, however the tail index regression estimation results from 1993 to 2016 can be obtained upon request). The tail index estimate in (\ref{alpha_t}), $\hat{\alpha}_{t}$, provides an estimate of the unconditional tail index
after considering the characteristics of all individuals in the sample for
each year.

The results in Table 3 show that in general, the estimates based on the method that ignores censored data underestimate the true effects of the variables. This is especially clear in the estimates for 2017 (e.g. female and finance). It is a consequence of the potential inconsistency of the uncensored estimates as highlighted in the Monte Carlo section above. However, the direction of the impact of the covariates suggested by the uncensored estimation is consistent with the results obtained from the censored tail index regression.

Comparing 1992 and 2017 (censored estimation), we generally observed a decrease in the estimates for 2017 (e.g. Transportation and Craft and Precision). In some cases, although in general not statistically significant (see e.g. other races, married no spouse, widowed, Transports and Trade), positive estimates in 1992 become negative in 2017 and vice versa. Considering only the statistically significant covariates it can be observed that Female, Black, Divorced, Single, Low Skill, Craft, Operators, Transportation and Public have a positive impact leading to a reduction in the probability of individuals with these characteristics being in the right tail, whereas Age, Education, Construction, Finance and Repair have a negative impact, originating an increase of the probability of individuals with these characteristics being in the right tail.

\begin{table}[htbp]
	\centering
	\def\sym#1{\ifmmode^{#1}\else\(^{#1}\)\fi}
	\caption{Uncensored and Censored Tail Index Regression Results\label{t3}}
	\sisetup{table-space-text-post = \sym{***}}
	\scalebox{0.80}{\begin{tabular}{ll*{4}{S[table-align-text-post=false]}}
			\toprule
			& & \multicolumn{2}{c}{Uncensored} & \multicolumn{2}{c}{Censored} \\
			& &\multicolumn{1}{c}{1992}  &  \multicolumn{1}{c}{2017}  &  \multicolumn{1}{c}{1992}  &  \multicolumn{1}{c}{2017} \\
			\midrule
			Constant & & 2.597 \sym{***} & 2.321 \sym{***} & 2.723 \sym{***} & 2.812 \sym{***} \\
			& & (0.066) &        (0.076) &        (0.075) &        (0.121)   \\
			Age &   & -0.011  \sym{***} & -0.007  \sym{***} & -0.013  \sym{***} & -0.012  \sym{***} \\
			& &( 0.001) &        (0.001) &        (0.001)        &( 0.001)   \\
			Female & & 0.272  \sym{***} & 0.134  \sym{***} & 0.309  \sym{***} & 0.229  \sym{***} \\
			& & ( 0.014) &        (0.013) &        (0.016) &       (0.022)   \\
			Education & & -0.076  \sym{***} & -0.061  \sym{***} & -0.085  \sym{***} & -0.100  \sym{***} \\
			& & (0.004) &     (0.004) &    (0.004) &    (0.006)   \\
			\multicolumn{2}{l}{\textbf{Race}}         &     &    &         \\
			&Black  & 0.085  \sym{***} & 0.004 &        0.098  \sym{***} & 0.010  \sym{***} \\
			& & ( 0.029) &        (0.034) &        (0.031) &        (0.054)   \\
			&Other race  & -0.023  \sym{***} & 0.004 &        -0.031        & 0.015   \\
			& & (0.025) &        (0.016) &        (0.030) &        (0.033)   \\
			\multicolumn{2}{l}{\textbf{Marital Status}}         &     &    &         \\
			&Married no spouse & 0.075        & -0.036 &        0.095 &        -0.100   \\
			& & (0.063) &        (0.046)        & (0.070)        & (0.091)   \\
			&Separated & 0.035        & 0.051        & 0.044        & 0.093   \\
			& & (0.042) &        (0.059) &        (0.047) &       (0.096)   \\
			&Divorced & 0.018        & 0.066  \sym{***} & 0.025        & 0.119 \sym{***} \\
			& & (0.019) &        (0.022)        & (0.022)        & (0.036)   \\
			&Widowed & 0.003        & -0.070        & 0.022        & -0.083   \\
			& & (0.083)   &      (0.052)        & (0.077)        & (0.099)   \\
			&Single & 0.009        & 0.066  \sym{***} & 0.015        & 0.114  \sym{***} \\
			& & (0.018) &        (0.018) &       (0.021) &       (0.029)   \\
			\multicolumn{2}{l}{\textbf{Occupations}}         &   &      &         \\
			&Administrative & 0.113  \sym{***} & 0.021        & 0.125  \sym{***} & 0.006   \\
			& & (0.020) &        (0.022)        & (0.023)        & (0.047)   \\
			&Low Skill & 0.098        & 0.093 &       0.111 \sym{*} & 0.153 \sym{*} \\
			& & (0.064) &        (0.060)        & (0.067)        & (0.091)   \\
			&Craft & 0.289  \sym{***} & 0.069 &        0.324  \sym{***} & 0.154  \sym{*} \\
			& & (0.033) &        (0.056) &        (0.035) &        (0.086)   \\
			&Operators & 0.394  \sym{***} & 0.148  \sym{***} & 0.416  \sym{***} & 0.217  \sym{***} \\
			& &(0.050)  &      (0.062) &        (0.053) &        (0.090)   \\
			&Transportation & 0.366  \sym{***} & 0.184  \sym{***} & 0.400 \sym{***} & 0.264 \sym{***} \\
			& & (0.023) &        (0.030) &        (0.025) &        (0.045)   \\
			\multicolumn{2}{l}{\textbf{Industry}}         &     &    &         \\
			&Construction & -0.202  \sym{***} & -0.117  \sym{***} & -0.234  \sym{***} & -0.173 \sym{***} \\
			& & (0.039) &        (0.042)        & (0.047) &        (0.071)   \\
			&Manufacturing & 0.003        & -0.011        & 0.004        & 0.026   \\
			&  &(0.028) &        (0.029)        & (0.031)        & (0.052)   \\
			&Transports & 0.050  \sym{***} & -0.016        & 0.051  \sym{**} & -0.001   \\
			& & (0.021) &        (0.027) &        (0.024) &       (0.040)   \\
			&Trade & 0.002 &        -0.052  \sym{**} & 0.001        & -0.052   \\
			&  & (0.020) &        (0.024) &        (0.025) &        (0.043)  \\
			&Finance & -0.126  \sym{***} & -0.099  \sym{***} & -0.184  \sym{***} & -0.219  \sym{***} \\
			& & (0.020) &        (0.022) &        (0.026) &        (0.043)   \\
			&Repair & -0.059  \sym{***} & -0.098  \sym{***} & -0.071  \sym{**} & -0.142  \sym{***} \\
			& & (0.028) &        (0.022) &        (0.033) &        (0.041)   \\
			&Personal & 0.125  \sym{***} & 0.048  \sym{***} & 0.127 \sym{***} & 0.060  \sym{*} \\
			& & (0.016) &       (0.018)        & (0.019)        & (0.033)   \\
			&Public & 0.238  \sym{***} & 0.120  \sym{***} & 0.272  \sym{***} & 0.265 \sym{***} \\
			& & (0.022)    & (0.024)        & (0.025)        & (0.040)   \\
			\bottomrule
	\end{tabular}}
	\begin{minipage}{15cm}
		\begin{spacing}{0.8}
			\bigskip
			\small{\textbf{	Note: }	This table reports the tail index regression results for 1992 and 2017. The first two columns present the uncensored results while the last two columns contain the censored results. The omitted categories are white, married, working as manager and the agriculture industry. Standard errors in parentheses and *, **, *** indicate significance at the 10\%, 5\% and 1\% significance levels.}
		\end{spacing}	
	\end{minipage}
\end{table}%

\FloatBarrier

\begin{figure}[h!]
	\caption{Uncensored and censored right tail index estimates from 1992 to 2017}
	\label{f2}
	\begin{center}
		$%
		\epsfxsize=4in \epsffile{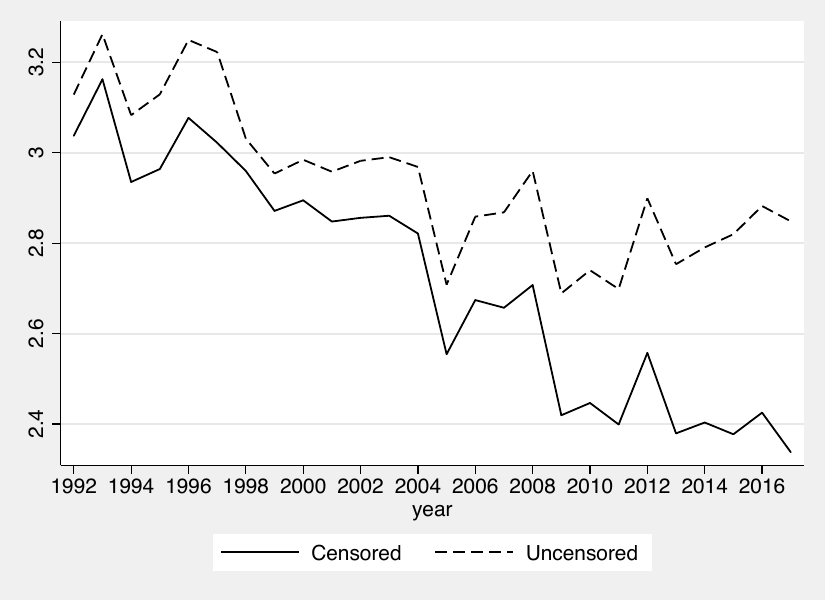}
		$%
		\bigskip
		\par
		\begin{minipage}{14cm}
			\begin{spacing}{0.8}
				\textbf{Note}: This figure reports the yearly right tail index estimates computed from both the uncensored and the censored regressions.
			\end{spacing}	
		\end{minipage}
	\end{center}
\end{figure}
\FloatBarrier

Figure \ref{f2} plots the tail index estimate computed as suggested in (\ref{alpha_t}) for each year from 1992 to 2017, based on uncensored and censored tail index regression estimates.  As discussed above, the uncensored approach misrepresents the true unconditional tail index,
as it ignores the top-coded wages, leading to the overestimation of the true values of $%
\alpha _{t}$ (where the bias/inconsistency is due to the fact that the top extreme
values are simply not included in the estimation). On the contrary, the censored
estimates take the information of the individuals of the top-coded wages, although the true wages are
not known, into account. Figure \ref{f2} shows that the censored estimates of $\hat{\alpha%
}_{t}$ have declined over the last 20 years. In other words, this Figure shows that the probability of
observing an extreme value today is higher compared to the 90s or even in the more recent past. This finding supports the idea that
upper-tail inequality has increased since the 90s and has become more
pronounced over the last 20 years. \citet{Autoretal(2008)} observed that the 90th percentile wage rose by more than 55$\%$ relative to the 10th percentile between 1963 and 2005, which represents a significant increase. However, our approach suggests that the increase in inequality found by \citet{Autoretal(2008)} may be a lower bound of the true increase in inequality.

Not adequately handling the top-coding may also lead to overestimation of the tail index in a variety of other applications (this has been recognized in e.g. the analysis of returns to education by \cite{Hubbard(2011)}, and differences in gender and race by \citet{BurkhauserLarrimore(2009a)} who already use approaches to accommodate for the fact that the wage data is top-coded). Our procedure provides additional flexibility by allowing researchers to evaluate which determinants impact the probability of being in the right tail of the wage distribution and which determinants do not. Thus, permits a more detailed analysis on how inequality has spread across industries, occupation, gender and other population characteristics.

\begin{figure}[thp]
	\caption{Partial Effects of Variables when u=0.15} \label{Fig5}
\begin{center}
	\includegraphics[height=5.0in, width=1 \textwidth]{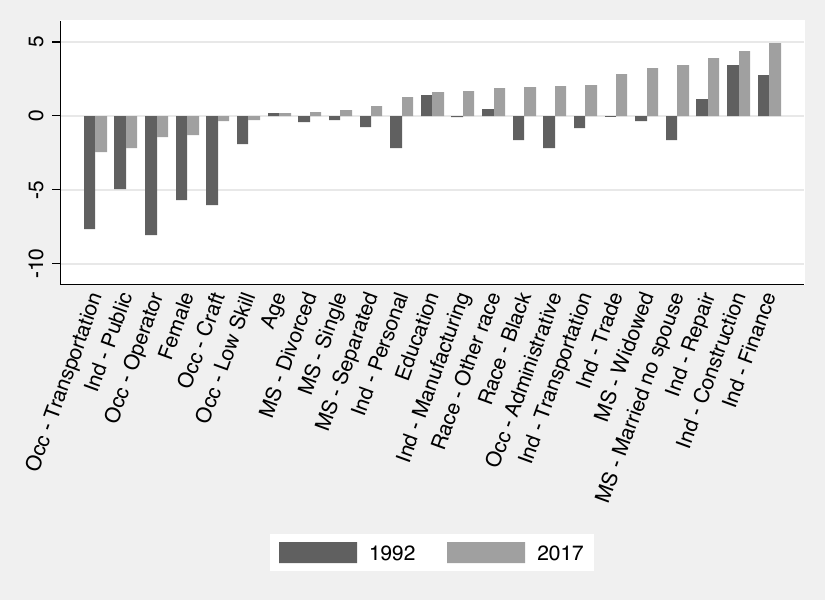}
\par
\bigskip
\begin{minipage}{16.5cm}
	\begin{spacing}{0.85}
	\small{	\textbf{Note: }For a description of the occupations see note under Table \ref{t2}.}
	\end{spacing}	
\end{minipage}
\end{center}
\end{figure}

In what follows, we focus on the
(conditional) tail index regression, and especially on the partial effects computed 
as discussed in section 2.2. Figure \ref{Fig5} groups the partial effect estimates and rearranges them from the lowest effect to the highest. A negative (positive) regression coefficient translates into a positive (negative) partial effect, which is associated with an increase (decrease) in the likelihood of
having more extreme values (the right tail becomes (less) heavier).

In the discussion that follows on the partial effects of the covariates, whenever we refer to an extreme event, we are referring to observations which are larger than the 96th quantile ($u=0.15$); see Remark 2 for details.

It is clear that industries such as Finance, Construction and Repair are the ones with more extreme wages. The probability of observing an extreme value increased in 2017 by 4.94\% for an individual working in the Finance industry. The probability of an individual working in the public industry having a wage higher than the 96th percentile decreased in 2017 by 2.14\%. The biggest increase from 1992 to 2017 was observed for individuals that are black, married without a spouse, or widowed. Older and more educated workers continued to have a significant probability of observing an extreme wage but there was no relevant change between 1992 and 2017.  Women observed a positive increase between 1992 and 2017, but the impact is still towards an increase in alpha (although smaller in absolute value than in 1992), i.e., a decrease in the probability of being in the right tail. In specific, women in 2017 have a partial effect on the tail index of -1.26\% which corresponds to a decrease in the probability of an extreme value. 

The impact in terms of occupation is very interesting. In comparison to an individual working in a non-routine cognitive occupation (managers) all other occupations display a positive contribution to observe an extreme value (although smaller in 2017). However, the picture is very different across occupations. Individuals working in routine occupations such as administrative workers reduced their presence in the right tail from 10\% to 8\% (see Table \ref{t2}) but the probability to observe an extreme value for these occupations increased in 2017 (-2.14\% in 1992 and changed to 2.00\% in 2017). The contribution to the probability of observing an extreme value for an individual working as an operator was significantly higher in 1992 than in 2017 (the partial effect was -8.04\% in 1992 and it reduced to -1.41\% in 2017).


\FloatBarrier

\subsection{Imputed mean wages}
\subsubsection{Imputing wages above the top-code}

A further important contribution of the approach introduced is its use for the imputation of mean wages above the top-code as described in Section 3.2. Some authors use values to adjust the top-coded wage which are time-varying and differ by group. For instance, \cite{MacphersonHirsch(1995)} provide separate Pareto estimates according to gender and by year from 1973 to 2014 using public CPS-Merged Outgoing Rotation Groups (CPS-MORG). These authors indicate that these values increase over time and are higher for men than for women (e.g. for 2014 the adjustment coefficient is 2.06 for men and 1.81 for women). In contrast, our analysis is based on the March CPS (outgoing rotation groups 4 and 8), and on the weekly wages instead of annual earnings, however, we also find evidence in favour of changing adjustment parameters.

This renders support to the observation that imputed wages above the top-code, based on a fixed value may lead to misstatement of results, given that this approach considers wages above the top-code to be independent of time, age, gender, race and other personal characteristics; as well as industry and occupation.

\begin{figure}[htp!] 
		\centering
		\includegraphics[width=0.7\textwidth]{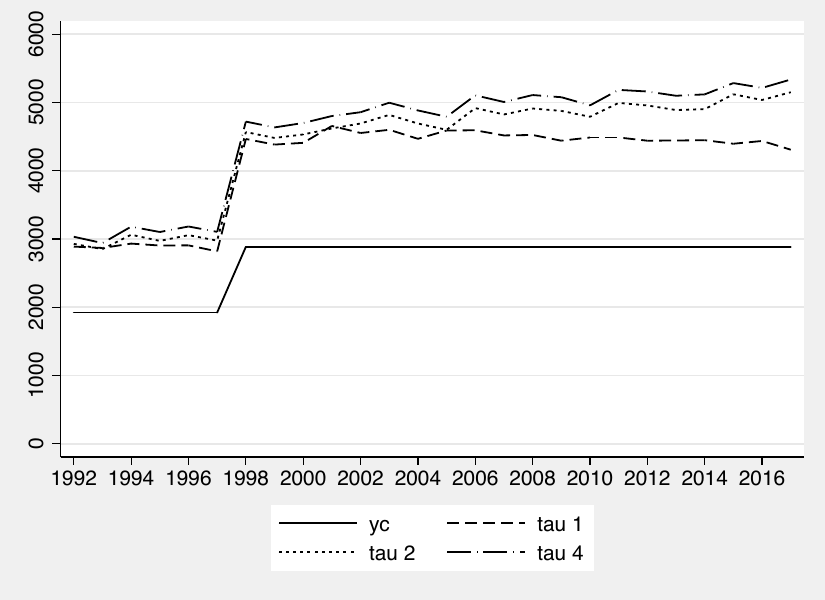}
		\caption{Prediction of Topcoded Wages} \label{FigPred1}
	\end{figure}

To impute wages above the top-code, we consider the estimate of  $%
E\left( \left. y_{i}\right\vert y_{i}>y_{c};\mathbf{x}_{i}\right) $
given by  $\hat{\tau}_{3}\left( \mathbf{x}_{i},y_{c}\right) $
in (\ref{tau3}). However, for some individuals in the CPS database,
especially those in the highest wage groups, $\hat{\alpha}%
\left( \mathbf{x}_{i}\right) $ can be in the neighborhood of 1, or even
lower than 1, which implies that $E\left( \left. y_{i}\right\vert
y_{i}>y_{c};\mathbf{x}_{i}\right) $ does not exist and, therefore, the
estimate $\hat{\tau}_{3}\left( \mathbf{x}_{i},y_{c}\right) $ is inadequate.
For these cases, we use the conditional median of $y_{i}$ given $y_{i}>y_{c}$
and $\mathbf{x}_{i},$ which is $2^{1/\hat{\alpha}\left( \mathbf{x}%
	_{i}\right) }y_{c}.$ In specific, to accommodate all situations (i.e. low and high values
of $\hat{\alpha}\left( \mathbf{x}_{i}\right) )$ we propose the following
estimator,
\begin{equation}
\hat{\tau}_{4}\left( \mathbf{x}_{i},y_{c}\right) =\left\{ 
\begin{array}{cc}
2^{1/\hat{\alpha}\left( \mathbf{x}_{i}\right) }y_{c} & 0<\hat{\alpha}\left( 
\mathbf{x}_{i}\right) \leq c \\ 
\frac{\hat{\alpha}\left( \mathbf{x}_{i}\right) }{\hat{\alpha}\left( \mathbf{x%
	}_{i}\right) -1}y_{c} & \hat{\alpha}\left( \mathbf{x}_{i}\right) >c%
\end{array}\right.. \label{median}
\end{equation}
In the empirical application of this statistic we set $c = 1.5$, since using  $c=1$ may lead to explosive estimates as $\frac{\hat{\alpha}\left( \mathbf{x}_{i}\right) }{\hat{\alpha}\left( \mathbf{x%
	}_{i}\right) -1}\rightarrow \infty $ as $\hat{\alpha}\left( \mathbf{x}%
_{i}\right) \rightarrow 1^{+}$. Other values of $c$ in the neighborhood of $%
c=1.5$ yield basically the same results. In the application to the CPS data  we observed that  for the overwhelming majority of estimates $\hat{\alpha}\left( \mathbf{x}_{i}\right) >1.5$. Thus, for most cases, the
estimate $\hat{\tau}_{4}\left( \mathbf{x}_{i},y_{c}\right) $ coincides with the second branch of (\ref{median}), which is the $\hat{\tau}_{3}\left( \mathbf{x}_{i},y_{c}\right) $ estimator in (3.3). Hence, we use $\hat{\tau}_{4}\left( \mathbf{x}_{i},y_{c}\right) $
to impute wages above the top-code over all individuals of the sample across time (see Figure \ref{FigPred1}).

Figure \ref{FigPred1} illustrates the estimates of the imputed wages above the top-code computed from the different approaches discussed in (\ref{tau1}), (\ref{tau2}) and (\ref{median}). The $\hat{\tau}_{2}$ and $\hat{\tau}_{4}$ estimates are similar.
	This result is expected given that for the overall analysis $\hat{\tau}_{4}$ is based on the values of $
	\hat{\alpha}_{t}$ computed as in (\ref{alpha_t}), which provides an estimate of the unconditional tail index after considering the characteristics of
	all individuals in the sample. However, in the case where estimates for a particular group, occupation or industry are considered, the $\hat{\tau}_{3}$ estimates will certainly be different from the $\hat{\tau}_{2}$ estimates (see next section).
	
\FloatBarrier

When the proportion of wages above the top-code is relatively small (as for example, from 1992 to 2005), the difference between $\hat{\tau}_{1},$ $	\hat{\tau}_{2}$ and $\hat{\tau}_{4}$ is relatively small; however, as more wages are located in the top-coded category (as for example, in the years following
	2007), the effect of censored data becomes stronger and the bias  (underestimation) produced by the	Hill estimator more pronounced ($\hat{\tau}_{1}\left( y_{c}\right) $).

\begin{figure}[hbt!]
		\centering
		\includegraphics[width=0.7\textwidth]{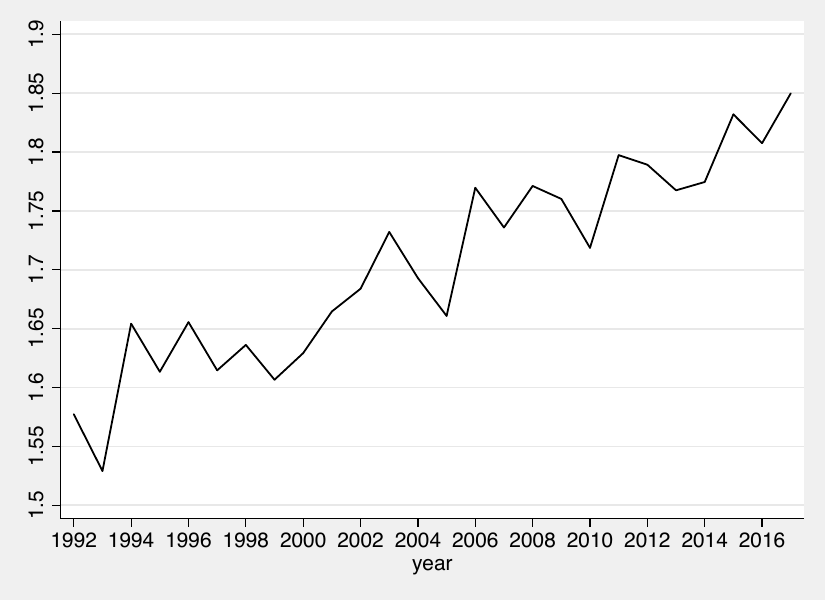}
		\caption{Top-coded wage adjustment factors between 1992 and 2017}  \label{FigPred2}
	\end{figure}

	Figure \ref{FigPred2} illustrates the time varying nature of the factor necessary to compute the imputed mean wages. Recall that to overcome the top-coding bias, in the literature, a constant value is frequently used to adjust the top-coded wages. For instance, \cite{AutorDorn(2013)} consider a sample between 1980-2005;  \cite{AcemogluAutor(2011)}, between 1973-2009; \cite{Autoretal(2008)} from 1963 to 2005;  \cite{KatzMurphy(1992)}	between 1963 and 1987; \cite{AutorDorn(2013)} between 1950 and 2005; \cite{Lemieux(2006)} from 1973 to 2003; and Beaudry, Green and Sand (2013) from 1979 to 2011.  Figure \ref{FigPred2} shows that using a fixed value may have been adequate for pre-1992 data,	but that the adjustment factor has increased over time reaching an overall value around 1.85 in 2017.

\FloatBarrier

\subsubsection{Imputed wages above the top-code by gender and industry}

\begin{figure}[h!]
	\caption{Imputed mean wages between 1992 and 2017 by industry}
	\label{f1aa}
	\begin{center}
		
		\scalebox{0.95}{		$
			\begin{array}{cc}
			\epsfxsize=3.1in \epsffile{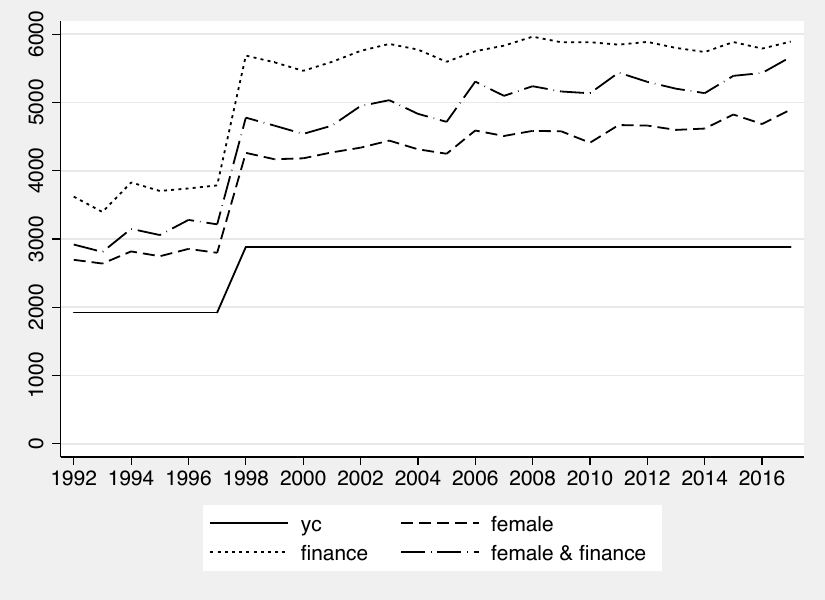}&  \epsfxsize=3.1in \epsffile{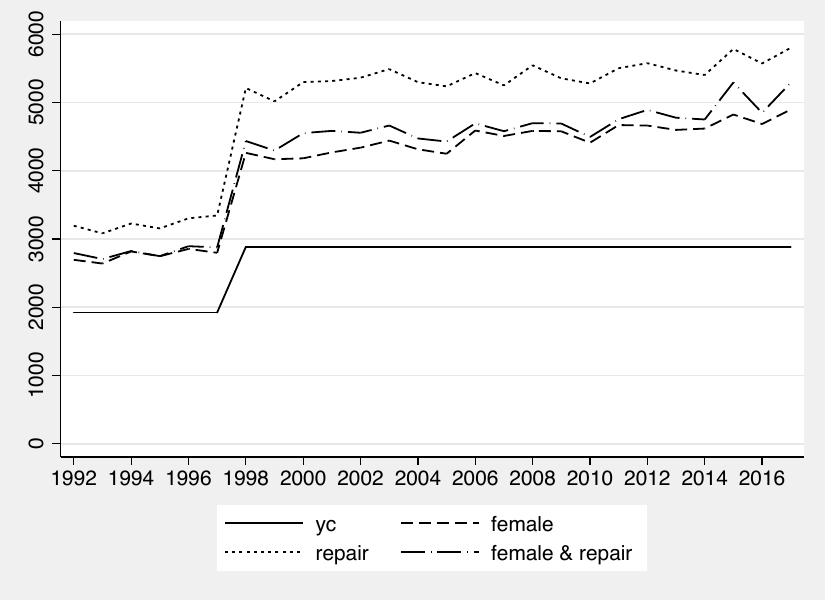}  \\
			\multicolumn{1}{c}{\mbox{\bf Ind: Finance} }&  \multicolumn{1}{c}{\mbox{\bf Ind: Repair}} \\%
			\epsfxsize=3.1in \epsffile{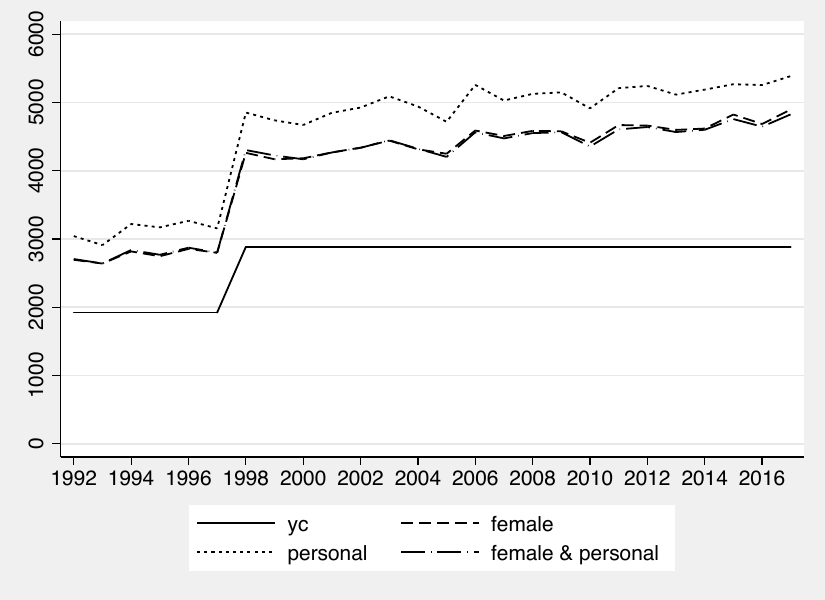} &\epsfxsize=3.1in \epsffile{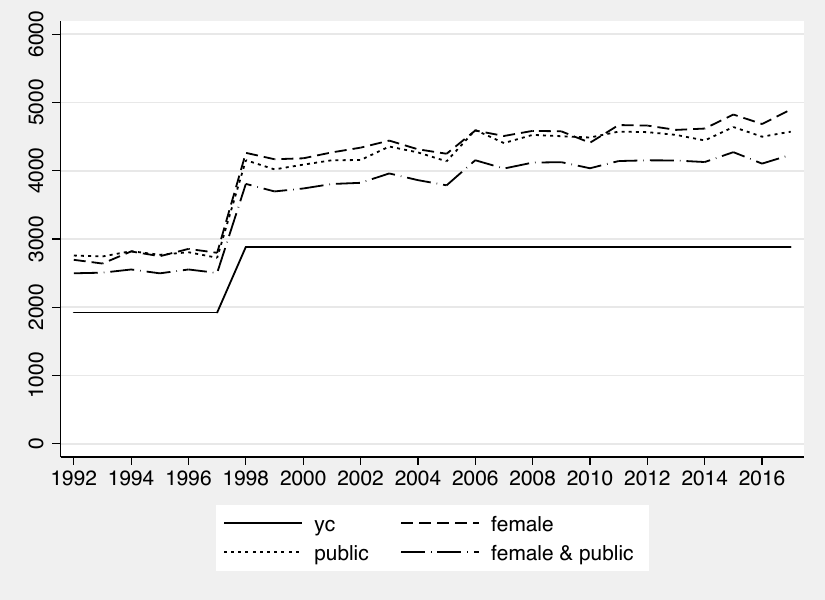}   \\
			\multicolumn{1}{c}{\mbox{\bf Ind: Personal}} &  \multicolumn{1}{c}{\mbox{\bf Ind: Public}}   \\%
			\end{array}%
			$}
	\end{center}
	
\end{figure}

Figure \ref{f1aa} illustrates the difference of the imputed mean values for individuals in the Finance, Repair, Personal and Public industries, as well as for women and women working in those industries. The purpose of these graphs is to further highlight the importance of allowing for different scaling factors depending on individuals characteristics and industry, but other graphs considering other characteristics can be plotted using our approach. 

The first noticeable result is that the imputed wage of individuals decreases when we compare the wages for individuals in the Finance, Repair, Personal and Public industries. Finance displays the largest and Public the lowest imputed wages of the four industries. With the exception of the Public industry, women's imputed wages are lower for the other three industries and this observation also holds when we condition women's imputed wages on the industry they are in. A further interesting result is that the imputed mean wages display an increasing trend over time in all industries, for women and for women in those industries, which is an indication that the adjustment factors used to compute the imputed wages also changes over time.

\FloatBarrier

This is further highlighted in Figure 10, where the graphs show the top-coded wage adjustment factors between 1992 and 2017, for different combinations of women working in different industries. Women’s top-coded adjustment factor is always smaller than the topcoded adjustment factor that would be applied to males in any industry with the exception of public. This implies that the public industry is the less heavy tailed. On top of that women working in the public industry earn less in the right tail than women in the right tail working in other industries.

While women working in personal and repair are not earning much more nor much less than in other industries we find that women working in finance would need a much higher adjustment factor. This means that this is the industry in which  they have been earning more and this result is reinforced with a clear positive trend between 1992 and 2017.

\begin{figure}[h!]
	\caption{Adjustment factors between 1992 and 2017 by industry to impute topcoded wages}
	\label{f10}
	\begin{center}
		
		\scalebox{0.95}{		$
			\begin{array}{cc}
			\epsfxsize=3.1in \epsffile{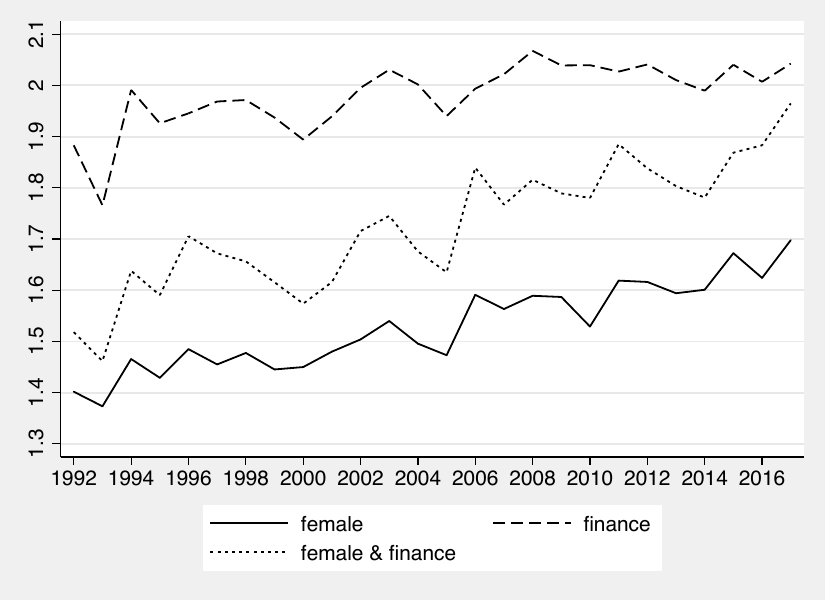}&  \epsfxsize=3.1in \epsffile{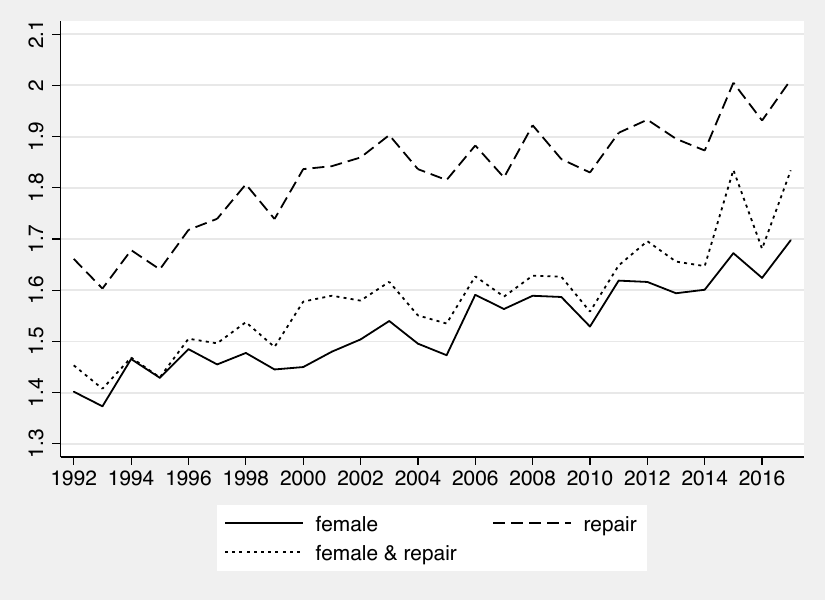}  \\
			\multicolumn{1}{c}{\mbox{\bf Ind: Finance} }&  \multicolumn{1}{c}{\mbox{\bf Ind: Repair}} \\%
			\epsfxsize=3.1in \epsffile{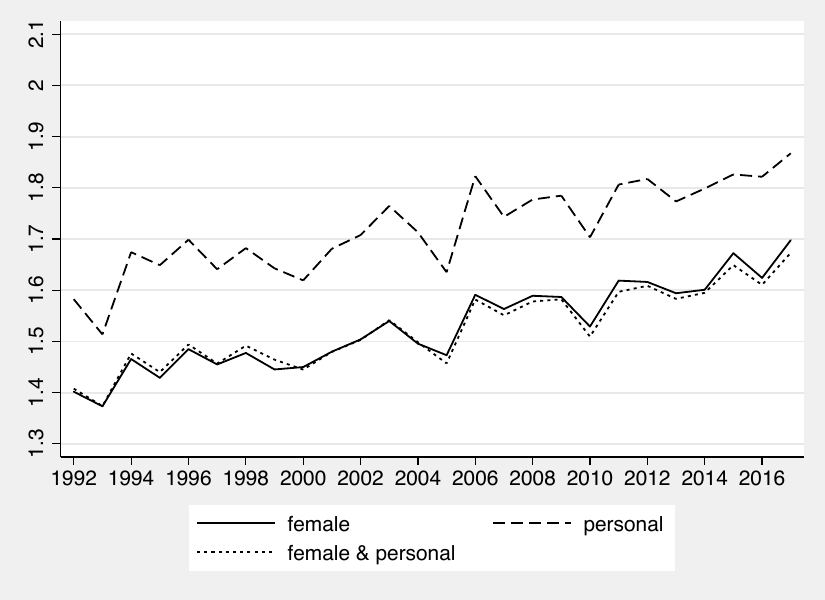} &\epsfxsize=3.1in \epsffile{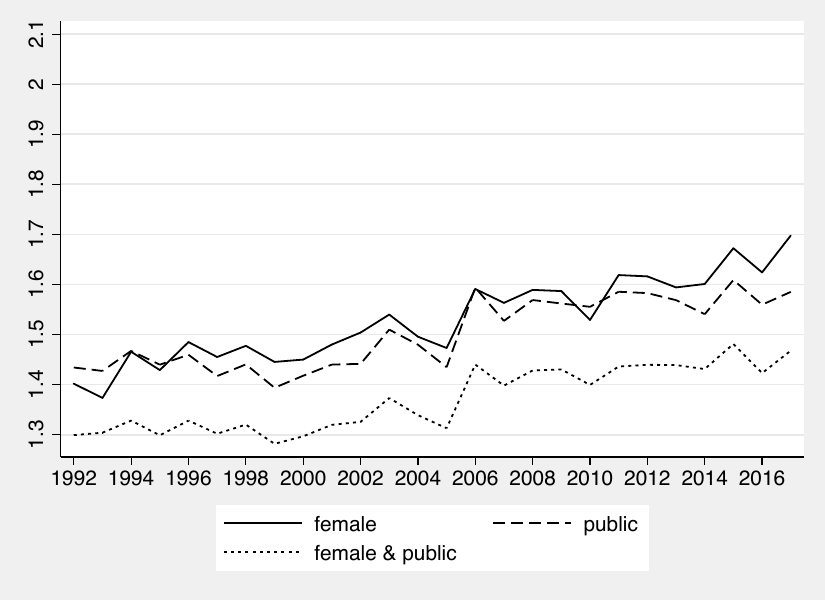}   \\
			\multicolumn{1}{c}{\mbox{\bf Ind: Personal}} &  \multicolumn{1}{c}{\mbox{\bf Ind: Public}}   \\%
			\end{array}%
			$}
	\end{center}
	
\end{figure}

\FloatBarrier

\newpage

\section{Conclusions}

This paper provides three important contributions to the literature. The first corresponds to the introduction  of a conditional tail index estimator which explicitly handles the top-coding problem and an indepth evaluation of its finite sample performance and comparison with competing methods. The Monte Carlo simulation exercise shows that the method proposed to estimate the tail index performs well in terms of estimation of the tail index and when used in the imputation of wages above the top-code when the sample is censored, which is an intrinsic feature of the public-use CPS database.

 Second, evidence is provided which shows that the factor values used to adjust the top-coded wages have changed over time and across the characteristics of individuals, occupations and industries and an indication of suitable values is proposed. Interestingly, the empirical results show that the upper-tail inequality has increased since the 90s and has become more pronounced over the last 20 years.
 
Third, an in depth empirical analysis of the dynamics of the US wage distribution's right tail using the public-use CPS database from 1992 to 2017 is provided. The application of the procedure to the CPS data reveals that individuals working  in industries such as finance, construction and repair are the ones with the more extreme wages. Moreover, it is also observed that the biggest increases in the probability of observing an extreme wage between 1992 and 2017 was for individuals that are black, married without a spouse, or widowed. Older and more educated workers continued to have a significant probability of observing an extreme wage but there was no relevant change between 1992 and 2017.  Women observed a positive increase between 1992 and 2017, but the impact is still towards an increase of alpha (although smaller in absolute values than in 1992) i.e. a decrease in the probability of an extreme wage. Furthermore, it is also noted that in comparison to an individual working in a non-routine cognitive occupation (managers) all occupations observed a positive contribution to observe an extreme value (although smaller in 2017). However, conclusions are different across occupations.

Our analysis also showed that women working in finance (public) would need a higher (lower) adjustment factor to impute the top-coded wages. Furthermore, we also observe that the adjustment factor used to impute top-coded wages should be adapted over time and across characteristics of the individuals especially when using censored data.

\bibliographystyle{apalike}
\bibliography{ref_ine}

\renewcommand{\thesection}{Appendix A:} \setcounter{equation}{0} \renewcommand{%
	\theequation}{A.\arabic{equation}}

\newpage

\section{Technical Appendix}

\noindent {\large{\textbf{A.1 Proof of Theorem 2.1} }}

For the proof of Theorem 2.1, let us show first that
$\mathbf{\Sigma}_{y_0}^{-1/2}\frac{\mathcal{\dot{K}}^{c}\left( \theta ,y_c\right) }{n}\overset%
{p}{\longrightarrow }0$. Consider that the sequence $\left\{ \left( y_{i},\mathbf{x}_{i}\right) \right\} $ is independently distributed.

\begin{eqnarray*}
	\mathbf{\Sigma}_{y_0}^{-1/2}\mathcal{\dot{K}}\left( \theta ,y_c\right)  &=&\sum_{t=1}^{n}\left\{ I
	_{\left\{ y_{0\leq }w_{i}<y_c\right\} }\left[ \exp \left( \mathbf{x}%
	_{i}^{\prime }\theta \right) \log \left( \frac{w_{i}}{y_{0}}\right)
	-1\right]\right.\\
	& &\left.-I_{\left\{ w_{i}=y_c\right\} }\exp \left( \mathbf{x}%
	_{i}^{\prime}\theta \right) \log \left( \frac{y_{0}}{y_c}\right) \right\}\mathbf{\Sigma}_{y_0}^{-1/2}\mathbf{x}_{i} \\
	&=&\sum_{t=1}^{n}e_{i}\mathbf{Z}_{ni}
\end{eqnarray*}%
where%
\begin{equation}
e_{i}=\left\{ 
\begin{array}{ll}
\exp \left( \mathbf{x}_{i}^{\prime }\theta \right) \log \left( \frac{%
	w_{i}}{y_{0}}\right) -1 ,\text{ for } & I_{\left\{ y_{0\leq
	}w_{i}<y_c\right\} } \\ 
-\exp \left( \mathbf{x}_{i}^{\prime }\theta \right) \log \left( \frac{y_{0}}{%
	y_c}\right),\text{ for } & I_{\left\{ w_{i}=y_c\right\} }%
\end{array}%
\right. \label{e_i}
\end{equation}
Thus, considering (\ref{e_i}) we cans how that,
\begin{eqnarray*}
	E\left( \left. e_{i}\right\vert \mathbf{x}_{i}\right) 
	&=&\int_{y_{0}}^{y_c}\left( \exp \left( \mathbf{x}_{i}^{\prime }\theta \right)
	\log \left( \frac{y}{y_{0}}\right) -1\right) f\left( \left. y\right\vert 
	\mathbf{x}_{i};\theta \right) dw-\exp \left( \mathbf{x}_{i}^{\prime }\theta
	\right) \log \left( \frac{y_{0}}{y_c}\right) P\left( I_{\left\{
		w_{i}=y_c\right\} }\right)  \\
	&=&\int_{y_{0}}^{y_c}\left( \exp \left( \mathbf{x}_{i}^{\prime }\theta \right)
	\log \left( \frac{y}{y_{0}}\right) -1\right) f\left( \left. y\right\vert 
	\mathbf{x};\theta \right) dw-\exp \left( \mathbf{x}_{i}^{\prime }\theta
	\right) \log \left( \frac{y_{0}}{y_c}\right) \left( 1-F\left( \left.
	y_c\right\vert \mathbf{x}_{i};\theta \right) \right)  \\
	&=&\left( \frac{y_{0}}{y_c}\right) ^{\exp \left( \mathbf{x}_{i}^{\prime
		}\theta \right) }\exp \left( \mathbf{x}_{i}^{\prime }\theta \right) \log
	\left( \frac{y_{0}}{y_c}\right) -\exp \left( \mathbf{x}_{i}^{\prime }\theta
	\right) \log \left( \frac{y_{0}}{y_c}\right) \left( \frac{y_{0}}{y_c}\right)
	^{\exp \left( \mathbf{x}_{i}^{\prime }\theta \right) } \\
	&=&0
\end{eqnarray*}

Moreover, it follows that 
\begin{eqnarray*}
	E\left( \left. e_{i}^2\right\vert \mathbf{x}_{i}\right) 
	&=&\int_{y_{0}}^{y_c}\left( \exp \left( \mathbf{x}_{i}^{\prime }\theta \right)
	\log \left( \frac{y}{y_{0}}\right) -1\right) f\left( \left. y\right\vert 
	\mathbf{x}_{i};\theta \right) dw-\exp \left( \mathbf{x}_{i}^{\prime }\theta
	\right) \log \left( \frac{y_{0}}{y_c}\right) P\left( I_{\left\{
		w_{i}=y_c\right\} }\right)  \\
	&=& 1-\left( \frac{y_{0}}{y_c}\right)^{\exp \left( \mathbf{x}_{i}^{\prime }\theta \right)}:=\Lambda.
\end{eqnarray*}

From Chebychev's weak law of law of large numbers, we have that%
\[
\mathbf{\Sigma}_{y_0}^{-1/2}\frac{\mathcal{\dot{K}}^c\left( \theta ,y_c\right) }{n}=\frac{1}{n}%
\sum_{t=1}^{n}e_{i}\mathbf{Z}_{ni}\overset{p}{\longrightarrow }E%
\left( e_{i}\mathbf{Z}_{ni}\right) 
\]%
and $E\left( e_{i}\mathbf{Z}_{ni}\right) $ is zero since%
\[
E\left( e_{i}\mathbf{Z}_{ni}\right) =E\left( E%
\left( \left. e_{i}\mathbf{Z}_{ni}\right\vert \mathbf{Z}_{ni}\right) \right) =%
E\left( E\left( \left. e_{i}\right\vert \mathbf{Z}%
_{ni}\right) \mathbf{Z}_{ni}\right) =0
\]

Given these results, considering that $\hat{\boldsymbol{\theta}}$ minimizes the log-likelihood function, $\frac{\mathcal{{K}}^c\left( \theta ,y_c\right) }{n}$ such that $\frac{\mathcal{\dot{K}}^c\left( \theta ,y_c\right) }{n}=0$, using the Mean Value Theorem it follows that
\begin{equation}
\frac{\mathcal{\dot{K}}^c\left( \theta ,y_c\right) }{n}=\frac{\mathcal{\dot{K}}^c\left( \theta_0, y_c\right) }{n}+\frac{\mathcal{\ddot{K}}^c\left( \theta_1,y_c\right) }{n}\left(\hat{\boldsymbol{\theta }}-\boldsymbol{\theta }_0\right)
\end{equation}
for some $\boldsymbol{\theta}_1 \in [\hat{\boldsymbol{\theta }}, \boldsymbol{\theta }_0] $. \hfill $\blacksquare$

\bigskip

\noindent {\large{\textbf{A.2 Computation of the partial effects}}}

We have%
\begin{eqnarray}
\delta &=&\frac{\bar{F}\left( \left. y\right\vert \Delta x+x\right) -\bar{F}
\left( \left. y\right\vert x\right) }{\bar{F}\left( \left. y\right\vert
x\right) }\times 100  \notag \\
&=&\frac{\left( \frac{y}{y_{0}}\right) ^{\alpha \left( \Delta x+x\right)
}-\left( \frac{y}{y_{0}}\right) ^{\alpha \left( x\right) }}{\left( \frac{y}{
y_{0}}\right) ^{\alpha \left( x\right) }}\times 100  \notag \\
&=&\left[\left( \frac{y}{y_{0}}\right)^{\alpha \left( \Delta x+x\right)-\alpha \left(x\right)
	}-1\right]\times 100 
\end{eqnarray}%
since $y=(1-u)^{\frac{1}{\alpha(x)}y_0}$ thus it follows that,

\begin{equation}
\delta =\left( \left( 1-u\right) ^{\frac{\alpha \left( \Delta x+x\right) }{\alpha
		\left( x\right) }-1}-1\right) \times 100.
\end{equation}%

Now and given the specification $\alpha \left( x\right) =\exp \left( \phi
\left( x\right) \right) $ where $\phi \left( x\right) $ is usually of type $%
\phi \left( x\right) =\mathbf{x}^{\prime }\mathbf{\beta }$ it follows that,
\begin{equation*}
\alpha \left( \Delta x+x\right) \simeq \alpha \left( x\right) +\frac{
d\alpha \left( x\right) }{dx}\Delta x =\alpha \left( x\right) +\phi ^{\prime }\left( x\right) \exp \left( \phi
\left( x\right) \right) \Delta x.
\end{equation*}%
Therefore%
\begin{equation*}
\frac{\alpha \left( \Delta x+x\right) }{\alpha \left( x\right) } =\frac{
\alpha \left( x\right) +\phi ^{\prime }\left( x\right) \exp \left( \phi
\left( x\right) \right) \Delta x}{\alpha \left( x\right) } =1+\phi ^{\prime }\left( x\right) \Delta x.
\end{equation*}%
In conclusion%
\begin{equation*}
\delta =\frac{\bar{F}\left( \left. y\right\vert \Delta x+x\right) -\bar{F}
\left( \left. y\right\vert x\right) }{\bar{F}\left( \left. y\right\vert
x\right) }\times 100 =\left( \left( 1-u\right) ^{\phi ^{\prime }\left( x\right) \Delta
x}-1\right) \times 100.
\end{equation*} \hfill $\blacksquare$

\newpage

\renewcommand{\thesection}{Appendix B:} \setcounter{equation}{0} \renewcommand{%
	\theequation}{B.\arabic{equation}}

\renewcommand\thefigure{B.\arabic{figure}}  
\setcounter{figure}{0}   
  
\section{Additional figures on proportion of individuals, according to characteristics, occupation and industry}
\begin{figure}[h!]
	\caption{Marital Status}
	\label{fA1}
	\begin{center}
		$%
		\begin{array}{cc}
		\epsfxsize=3.1in \epsffile{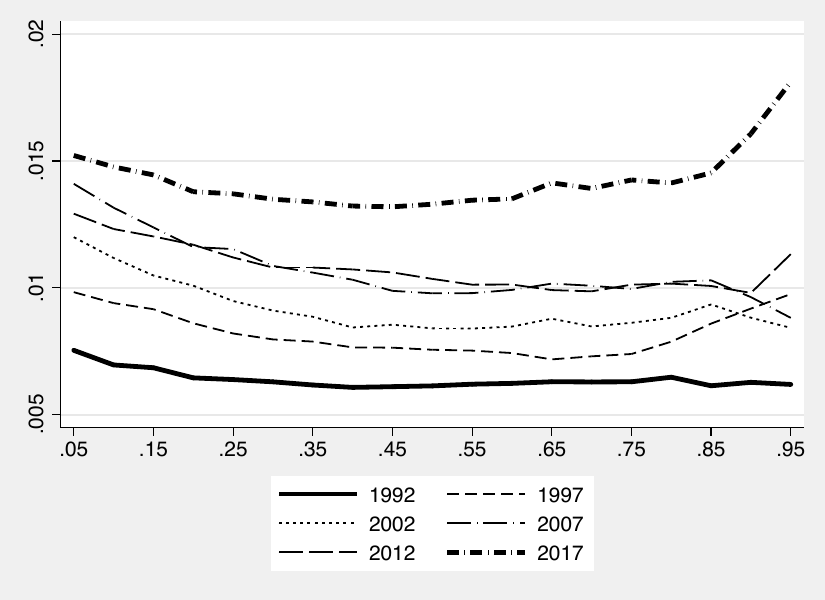} &  \epsfxsize=3.1in \epsffile{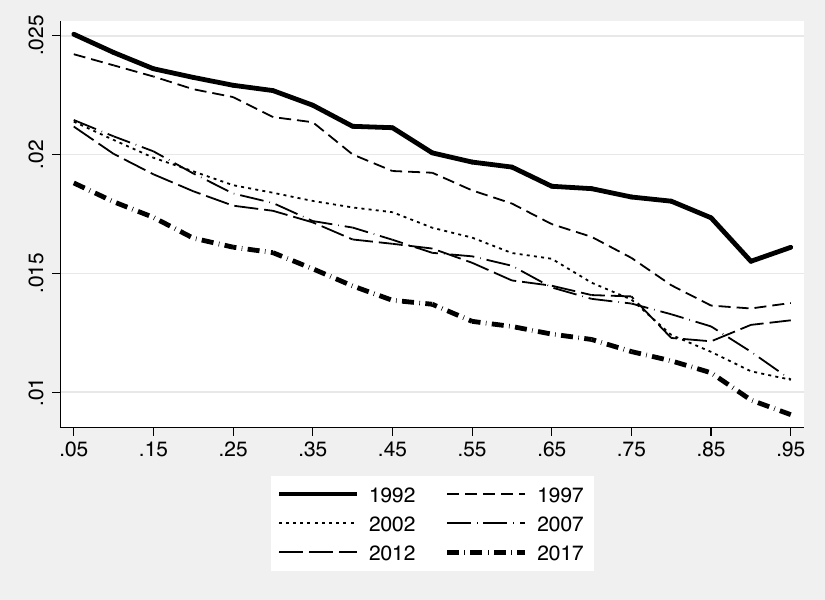}  \\
		\multicolumn{1}{c}{\mbox{\bf Married No Spouse}} &  \multicolumn{1}{c}{\mbox{\bf Separated}} \\%
		\epsfxsize=3.1in \epsffile{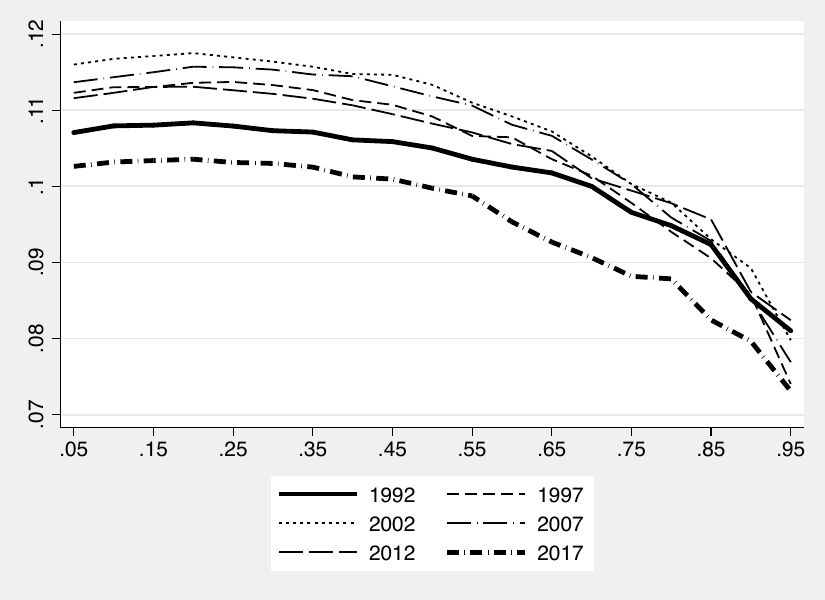} &  \epsfxsize=3.1in \epsffile{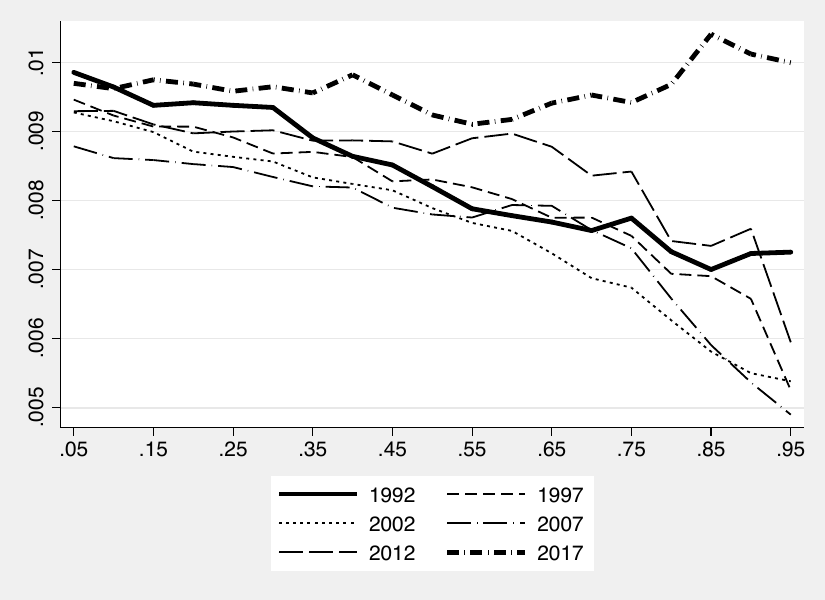}  \\
		\multicolumn{1}{c}{\mbox{\bf Divorced}} &  \multicolumn{1}{c}{\mbox{\bf Widowed}} \\%
		\end{array}%
		$
	\end{center}
\end{figure}
\FloatBarrier

\begin{figure}[bpth!]
	\caption{Occupation}
	\label{fA2}
	\begin{center}
		$%
		\begin{array}{cc}
		\epsfxsize=3.1in \epsffile{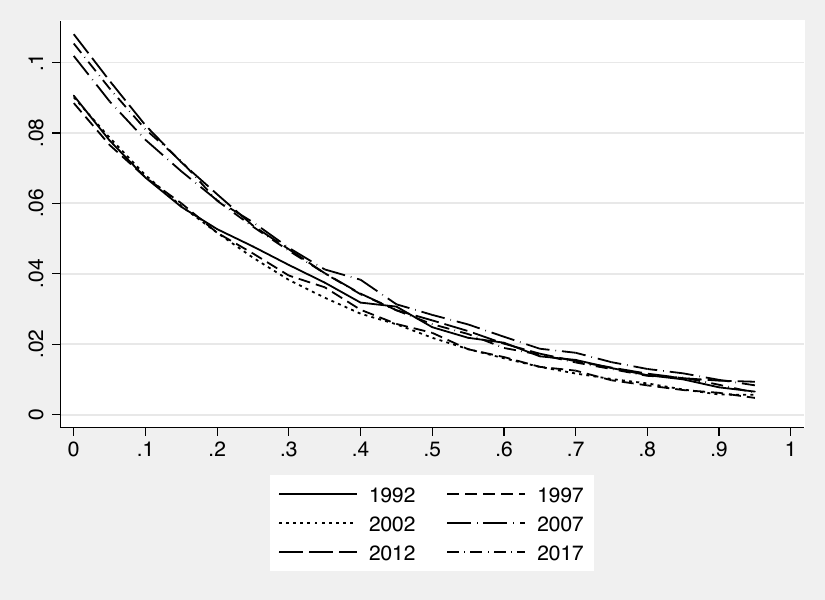} &  \epsfxsize=3.1in \epsffile{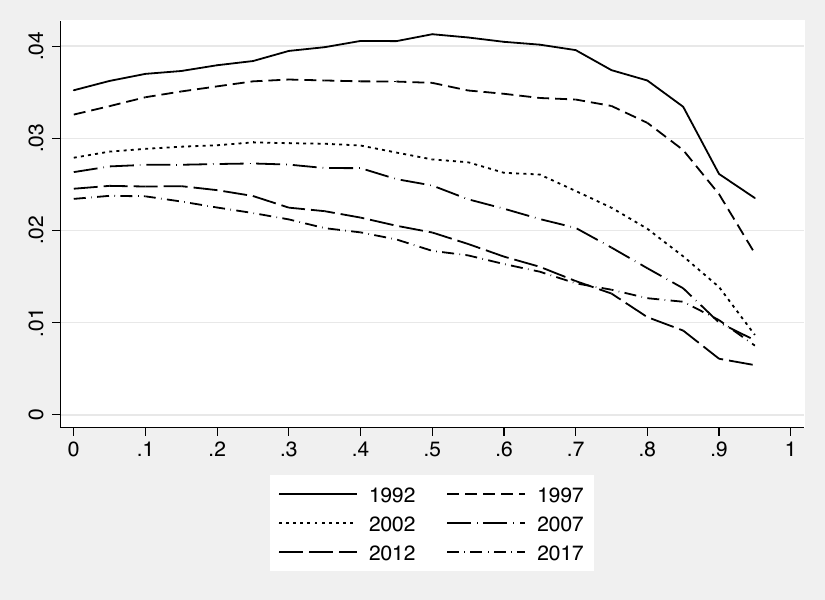}\\
		\multicolumn{1}{c}{\mbox{\bf Low Skill}} & \multicolumn{1}{c}{\mbox{\bf Craft}} \\%
		\epsfxsize=3.1in \epsffile{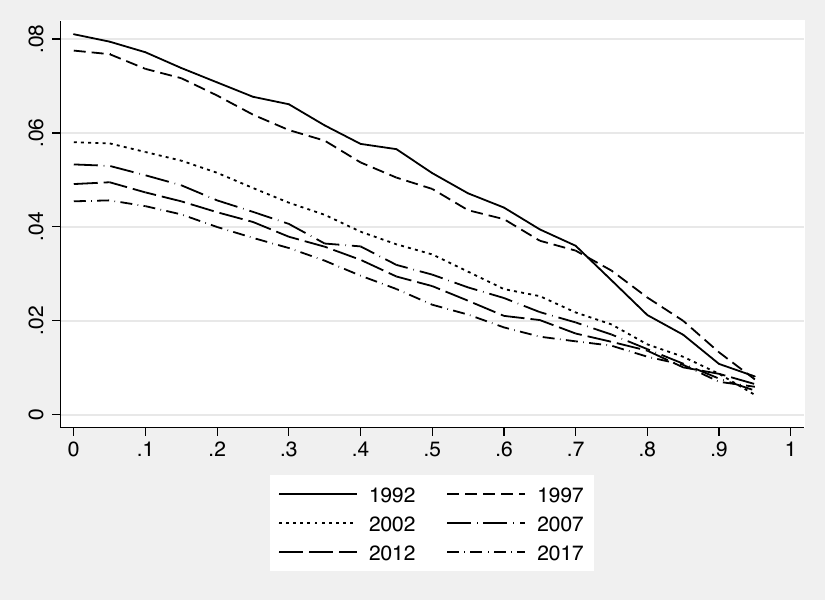} & \epsfxsize=3.1in \epsffile{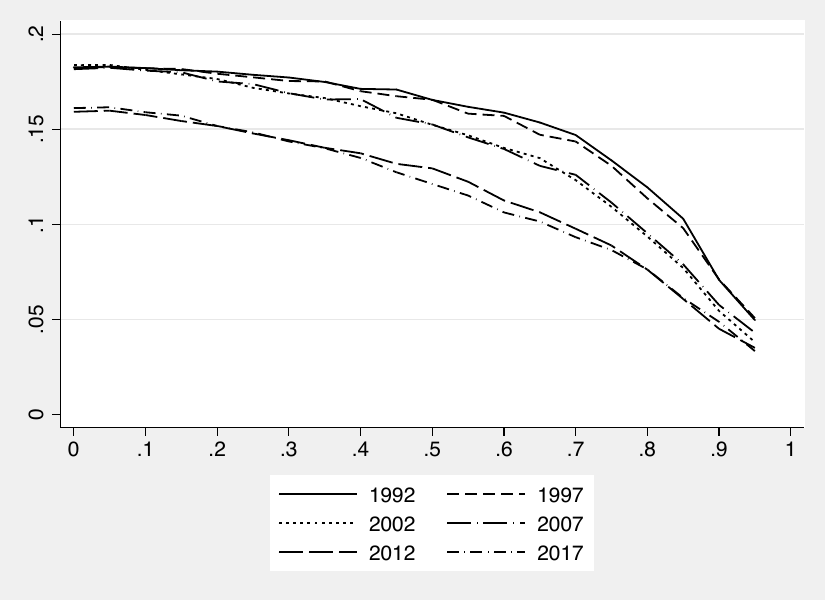}  \\
		\multicolumn{1}{c}{\mbox{\bf Operators}} &	\multicolumn{1}{c}{\mbox{\bf Transports}}\\
		\end{array}%
		$
	\end{center}
\end{figure}
\FloatBarrier

\begin{figure}[hbtp!]
	\caption{Industry}
	\label{fA3}
	\begin{center}
		$%
		\begin{array}{cc}
		\epsfxsize=3.1in \epsffile{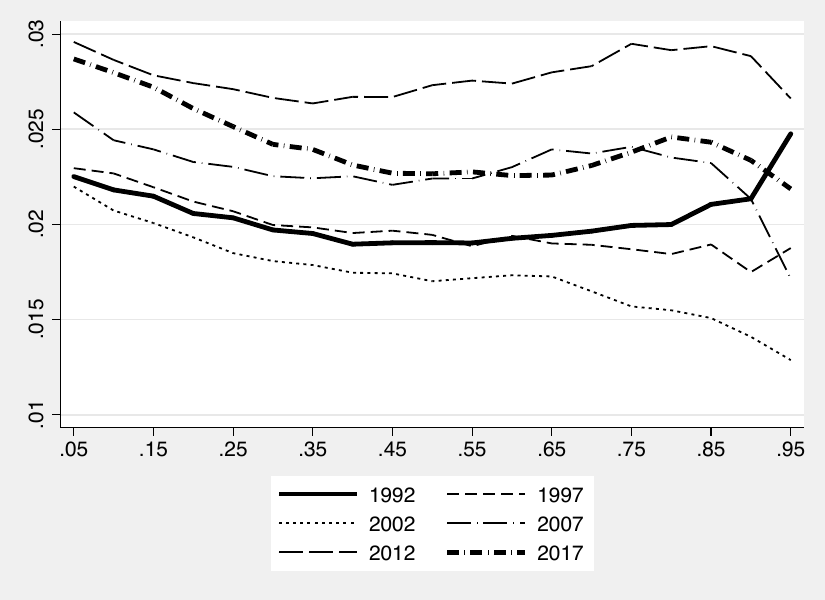} &  \epsfxsize=3.1in \epsffile{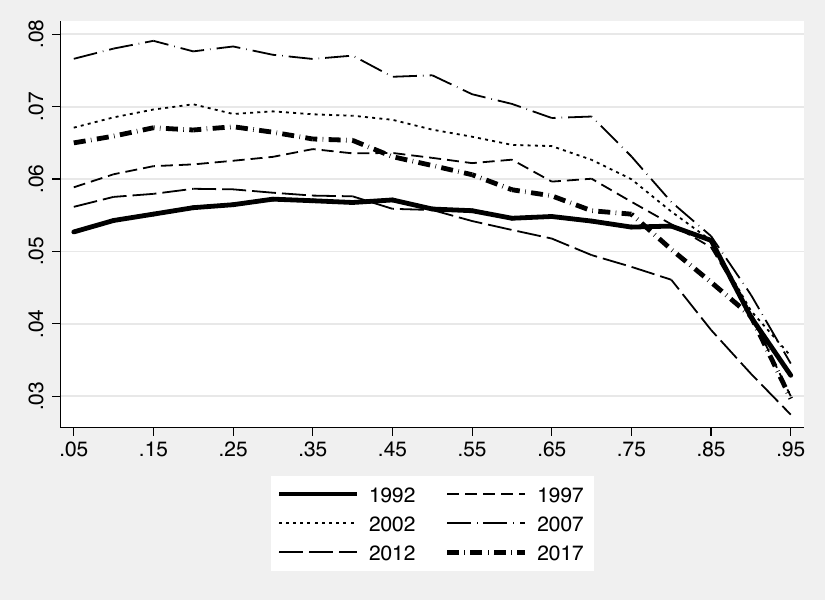}  \\
		\multicolumn{1}{c}{\mbox{\bf Agriculture}} &  \multicolumn{1}{c}{\mbox{\bf Construction}} \\%
		\epsfxsize=3.1in \epsffile{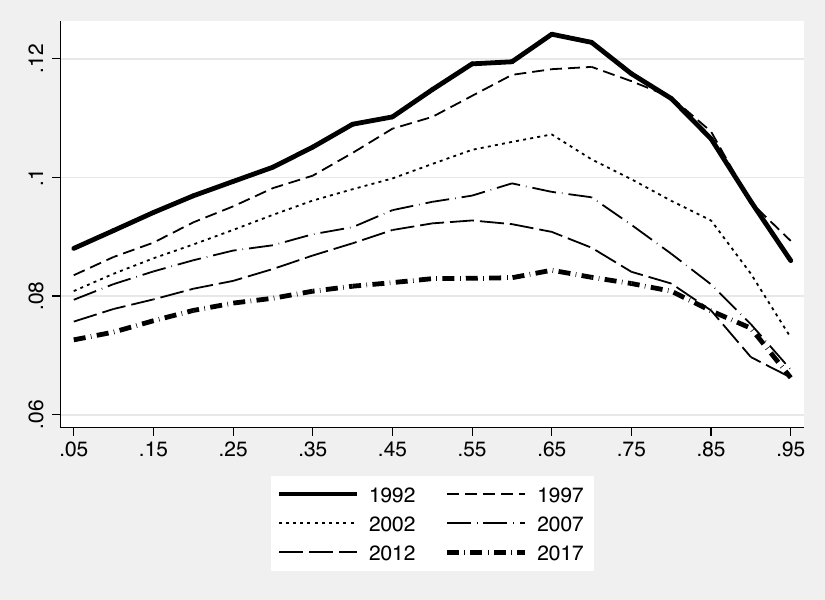} &   \epsfxsize=3.1in \epsffile{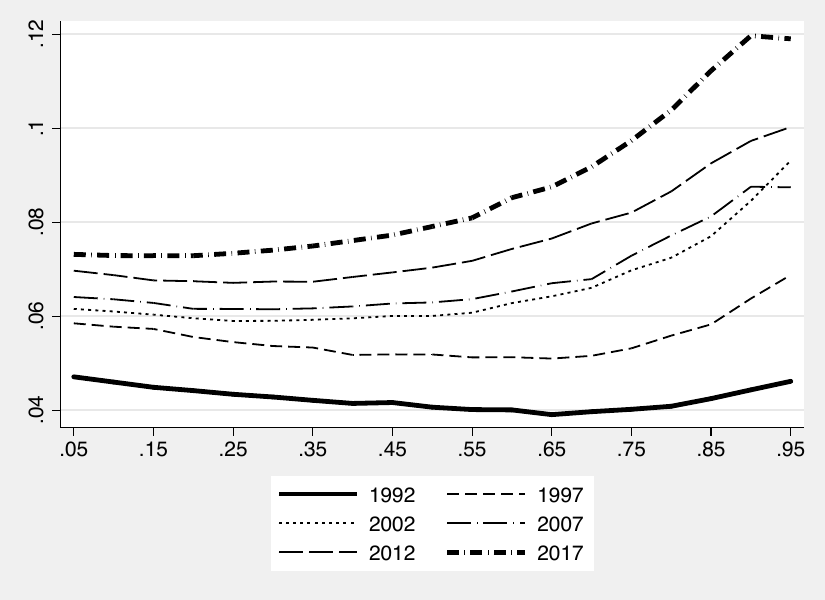} \\
		\multicolumn{1}{c}{\mbox{\bf Transport}} &  \multicolumn{1}{c}{\mbox{\bf Repair}}\\%
		\epsfxsize=3.1in \epsffile{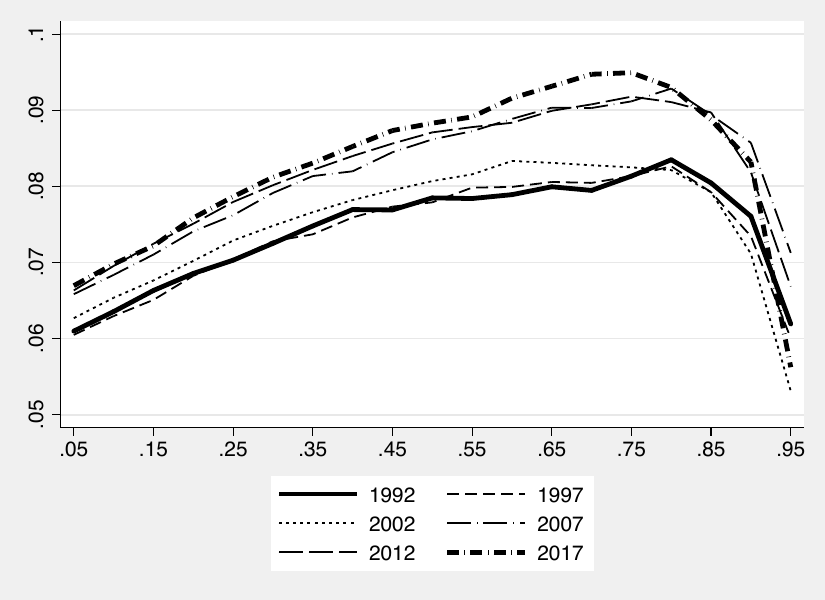} & \\
		\multicolumn{1}{c}{\mbox{\bf Public}} & \\%
		\end{array}%
		$
	\end{center}
\end{figure}

\FloatBarrier

\renewcommand{\thesection}{Appendix C:} \setcounter{equation}{0} \renewcommand{%
	\theequation}{C.\arabic{equation}}

\section{Industry classification} 

	\vspace{2.5cm}

\begin{center}
Table C.1 - Industry classification
	\vspace{0.25cm}
	
\begin{tabular}{p{0.2\textwidth}p{0.75\textwidth}}
\hline
\\
\textbf{Agriculture}& agriculture, forestry, fishing, hunting, mining and utilities; \\
\textbf{Construction}& only construction;\\
\textbf{Manufacturing}& manufacturing of non-durable and durable goods and wood;\\
\textbf{Transports}& transportation, warehousing, utilities electric light;\\
\textbf{Trade}& wholesale and retail trade;\\
\textbf{Finance}& finance and insurance;\\
\textbf{Repair}& business and repair;\\
\textbf{Personal}& personal services, entertainment and recreation, professional and related services;\\
\textbf{Public}& public administration and armed forces.\\
\\
\hline
\end{tabular}
\end{center}
\noindent \textbf{Note:} This classification is based on the harmonized variable "ind1990" from IPUMS.

\end{document}